\documentclass[showkeys,showpacs,superscriptaddress,twocolumn,aps,pre,10pt,nofootinbib]{iopart}

\usepackage{amssymb}
\usepackage{graphicx}
\usepackage[caption=false]{subfig}
\usepackage{mathtools}
\usepackage{cite}
\usepackage{units}
\usepackage[colorlinks=true, allcolors=blue]{hyperref}
\usepackage{xcolor}
\usepackage{cuted}
\usepackage[nolist]{acronym}
\usepackage{amsmath}
\usepackage{comment}

\begin{document}


\title{A First-Principles Closure for Nonlocal Magnetized Transport}


\author{N T Mitchell$^1$, D A Chapman$^2$, G Kagan$^1$}

\address{$^1$ The Blackett Laboratory, Imperial College, London SW7 2AZ, UK}

\address{$^2$ First Light Fusion Ltd., Unit 9/10 Oxford Pioneer Park, Mead Road, Yarnton, Kidlington OX5 1QU, UK}

\ead{\mailto{n.mitchell22@imperial.ac.uk}}


\begin{abstract}
    A reduced kinetic method (RKM) for describing nonlocal transport in magnetized plasmas is derived from first principles and considered in a 1D3V geometry. Unlike standard nonlocal closures, this RKM uses the Fokker-Planck collision operator, therefore local transport results are naturally reproduced for small Knudsen number. An inhibited peak heat flux and preheat of the conductive heat flux are observed, which are expected from physical arguments and previous kinetic studies. Nonlocal behavior of other transport fluxes, namely the Righi-Leduc, Peltier, Ettingshausen, Nernst, thermal force, friction, cross friction, viscous stress, and gyroviscous stress terms are also demonstrated. Neglecting the nonlinear component of the Fokker-Planck collision operator is justified \textit{a posteriori}. An especially computationally efficient and analytically simpler version of the RKM is presented.
\end{abstract}




\maketitle
\ioptwocol

\allowdisplaybreaks


\begin{acronym}
\acro{VFP}{Vlasov-Fokker-Planck}
\acro{MFP}{mean free path}
\acro{BGK}{Bhatnagar–Gross–Krook}
\acro{DF}{distribution function}
\acro{DDF}{deviation of distribution function from Maxwellian}
\acro{CoM}{centre-of-momentum}
\acro{RKM}{reduced kinetic method}
\acro{LTE}{local thermal equilibrium}
\acro{ICF}{inertial confinement fusion}
\acro{PIC}{particle-in-cell}
\end{acronym}
\section{Introduction}\label{section_intro}

Collisional transport of heat and momentum is an essential piece of physics for understanding and modeling of the collective behavior of plasmas, and therefore of great practical significance for confined fusion schemes \cite{J_D_Callen_1997,icf_kinetics_review}. If the plasma is sufficiently collisional, fluid equations and local closures for transport fluxes derived from the Chapman-Enskog expansion \cite{chapman_mathematical_1939,Ferziger_book} are justified. Here, `sufficiently' collisional requires that the \acp{MFP} of particles relevant to heat and momentum transport are shorter than typical length scales $L$ over which fluid quantities such as density and temperature vary, i.e. $\lambda_{\text{th}} / L \ll 1$, where $\lambda_{\text{th}}$ is the thermal \ac{MFP}. In addition, the collision time is required to be shorter than characteristic hydrodynamic timescales.

If the plasma is not sufficiently collisional, analytic local transport results become invalid. Since the \ac{MFP} of a single particle in an ideal plasma has a strong scaling with particle speed $\lambda\propto v^{4}$, suprathermal particles in the tail of the velocity distribution that contribute significantly to heat and momentum transport are far less collisional than the bulk. These tail particles may therefore stream across temperature gradients before they collide, depositing their energy and momentum far away in space. This nonlocal transport can significantly influence the conductive heat flux, suppressing the peak heat flow and preheating colder regions, both of which directly impact hydrodynamic evolution and therefore fusion yield. Nonlocal effects are particularly prevalent at low densities, high temperatures, and sharp gradients, all of which are often encountered in confined fusion plasmas. Understanding and appropriately modeling nonlocal heat transport is therefore an important consideration for hydrodynamic simulations to accurately simulate, design, and optimize confined fusion schemes. 

Kinetic effects such as nonlocal heat flow are not accounted for in a standard fluid framework. Full kinetic approaches such as \ac{VFP} and \ac{PIC} simulations naturally include these kinetic effects. However, full kinetic codes are often computationally intractable for large-scale integrated simulations of fusion-relevant scenarios such as \ac{ICF} implosions, often leaving fluid simulations as the preferred option. An intermediate approach is a fluid simulation with a nonlocal transport closure which provides transport fluxes to the fluid model for a given set of hydrodynamic profiles accounting for nonlocal effects. This hybrid approach then includes nonlocal transport effects while maintaining computational tractability. Developing an accurate nonlocal closure requires a thorough investigation of transport in nonlocal scenarios.


%
Previous literature is rich in nonlocal transport studies and closures that focus almost entirely on the electron conductive heat flux \cite{Epperlein91,nicolai_practical_2006,article,brodrick_testing_2017,arran_accuracy_2025,holec_nonlocal_2018,marocchino_comparison_2013,10.1063/1.4986095,lezhnin_particle--cell_2025,dearling_transport_2024}, neglecting the suite of other transport fluxes which may be influenced by nonlocal behavior. These closures almost always rely on simplified model collision operators and diffusive approximations, therefore do not naturally reproduce local transport results in the short mean free path limit and likely miss some parts of nonlocal physics and behavior. These approaches also overlook current-driven transport which appears naturally in local analytic closures. Currents generate magnetic fields, therefore a closure including current-driven transport must include magnetic field effects. 

This work presents a \ac{RKM} extended from previous work \cite{Mitchell_2024} to act as a closure to nonlocal magnetized transport with flows. This previous work considered a planar geometry with a single preferred direction (1D) in the absence of a magnetic field. This system has a cylindrical symmetry in the velocity space, therefore the velocity space became two dimensional (2V). Including a magnetic field component perpendicular to the single spatial direction breaks this cylindrical symmetry since there will be a net Lorentz force on particles acting in the $\boldsymbol{v}\times\boldsymbol{B}$ direction, which is typically not parallel to the single spatial direction. The addition of magnetic fields therefore requires three velocity dimensions, leaving the 1D2V geometry that previous work considered insufficient. Extending the \ac{RKM} to 1D3V requires expanding the \ac{DF} in spherical harmonics with complex-valued $m \neq 0$ modes, as oppsosed to the previous 1D2V implementation which expanded in Legendre polynomials with real-valued $m=0$ modes. The \ac{RKM} describes all classical transport quantities for both electrons and ions, including conductive heat flux, Righi-Leduc heat flux, Peltier heat flux, Ettingshausen heat flux, thermal force, Nernst force, friction, cross friction, and viscous stress, and gyroviscous stress terms. Standard nonlocal effects due to short length scales on the whole suite of transport fluxes are found, as well as novel nonlocal behavior due to fast flows on the Ettingshausen heat flow \cite{mitchell_nonlocal_2025}. The \ac{RKM} is derived from first principles with the Fokker-Planck collision operator, therefore it naturally reproduces local transport results to high accuracy across the full range of ionizations and magnetizations. Transport quantities are solved self-consistently, in particular the electron-ion momentum exchange.

The remainder of this paper is organized as follows. Section \ref{sec:local_results} provides a summary of local analytic transport results for electrons and ions in a simple plasma. Section \ref{section_model} presents and derives the \ac{RKM} for electrons starting from the \ac{VFP} equation and describes its numerical implementation. Section \ref{section_electron_results} applies the \ac{RKM} to investigate nonlocal behavior of both temperature-gradient-driven and current-driven transport terms in unmagnetized, moderately magnetized, and strongly magnetized cases. Section \ref{sec:kinetics} presents the underlying behavior of the \ac{DF} and heat flux integrand subject to magnetization and nonlocality. The validity of linearizing the Fokker-Planck operator in nonlocal regimes, a key assumption of the \ac{RKM}, is checked \textit{a posteriori}. A more tractable version of the \ac{RKM} that allows the field-particle operator to be dropped is also introduced. Section \ref{sec:ion_results} presents the \ac{RKM} for a single ion species and presents nonlocal behavior of the ion viscous stress tensor in an unmagnetized and magnetized case. Section \ref{section_conclusion} discusses these results. 

\section{Summary of local transport results}\label{sec:local_results}

Local closures for collisional transport in plasmas have been thoroughly investigated in previous literature \cite{Spitzer1962,Braginskii1965ReviewsOP,10.1063/1.865901,PhysRevLett.126.075001,10.1063/1.4801022,helander_and_sigmar}. This work compares against the transport coefficients provided by Ji and Held \cite{10.1063/1.2977983,ji_closure_2013,10.1063/1.4922755} which demonstrate good convergence.


\subsection{Electron transport}

Assuming a sufficiently collisional plasma, local results for the electron heat flux $\boldsymbol{q}_e^{\text{local}} = \boldsymbol{q}_e^T + \boldsymbol{q}_e^u$ and electron-ion momentum exchange $\boldsymbol{R}_e^{\text{local}} = \boldsymbol{R}_e^T + \boldsymbol{R}_e^u$ may be expressed as
\begin{align}
    \boldsymbol{q}_e^T &= - \kappa_c^e \big( (\underbrace{\hat{\kappa}^e_\parallel \boldsymbol{b}\boldsymbol{b}
    + \hat{\kappa}^e_\perp(\mathbb{I}-\boldsymbol{b}\boldsymbol{b}))\cdot\boldsymbol{\nabla}T_e }_{\text{Heat conduction}}
    + \underbrace{\hat{\kappa}^e_\times \boldsymbol{b}\times \boldsymbol{\nabla}T_e}_{\text{Righi-Leduc}} \big),
    \notag\\[0.5em]
    \boldsymbol{q}_e^u &= n_e T_e \big( 
    \underbrace{ (\hat{\beta}_\parallel \boldsymbol{b}\boldsymbol{b}
    + \hat{\beta}_\perp(\mathbb{I}-\boldsymbol{b}\boldsymbol{b}))\cdot \Delta\boldsymbol{u} }_{\text{Peltier}}
    + \underbrace{\hat{\beta}_\times \boldsymbol{b}\times \Delta\boldsymbol{u}}_{\text{Ettingshausen}} \big),
    \notag\\[0.5em]
    \boldsymbol{R}_e^T &= - n_e \big(
    \underbrace{ (\hat{\beta}_\parallel \boldsymbol{b}\boldsymbol{b}
    + \hat{\beta}_\perp(\mathbb{I}-\boldsymbol{b}\boldsymbol{b}))\cdot\boldsymbol{\nabla}T_e }_{\text{Thermal force}}
    + \underbrace{ \hat{\beta}_\times \boldsymbol{b}\times \boldsymbol{\nabla}T_e }_{\text{Nernst}} \big),
    \notag\\[0.5em]
    \boldsymbol{R}_e^u &= - \frac{n_e m_e}{\tau_{ee}} \big(
    \underbrace{ (\hat{\alpha}_\parallel \boldsymbol{b}\boldsymbol{b}
    + \hat{\alpha}_\perp(\mathbb{I}-\boldsymbol{b}\boldsymbol{b}))\cdot \Delta\boldsymbol{u} }_{ \text{Friction}}
    - \underbrace{ \hat{\alpha}_\times \boldsymbol{b}\times \Delta\boldsymbol{u} }_{\text{Cross friction}} \big),
\end{align}
where $\boldsymbol{b} = \boldsymbol{B}/  \vert \boldsymbol{B}\vert$ is the unit vector parallel to the magnetic field, and $\hat{\alpha}_{(\parallel,\perp,\times)}$, $\hat{\beta}_{(\parallel,\perp,\times)}$, and $\hat{\kappa}_{(\parallel,\perp,\times)}^e$ are dimensionless coefficients that depend only on the effective ionization $Z$ and, for perpendicular $\perp$ and cross $\times$ terms, the Hall parameter $\chi_e = \omega_{\text{c},e}\tau_{ee}$. The collision time and basic thermal conductivity are defined as $\tau_{\alpha\beta} = \frac{3\sqrt{\pi}}{4\hat{\nu}_{\alpha\beta}} = \frac{6\sqrt{2}\pi^{3/2}\varepsilon_0^2 m_\alpha^{1/2} T_\alpha^{3/2}}{n_\beta q_\alpha^2 q_\beta^2 \ln \Lambda_{\alpha\beta} } $ and $\kappa_c^\alpha = n_\alpha T_\alpha \tau_{\alpha\alpha} / m_\alpha$ respectively. The electron and ion flow velocities are $\boldsymbol{u}_e$ and $\boldsymbol{u}_i$ respectively, with the relative motion of electrons against ions $\Delta \boldsymbol{u} \equiv \boldsymbol{u}_e - \boldsymbol{u}_i = - \boldsymbol{j}/en_e$.

\subsection{Ion transport}

Ion transport has further complications compared to electron transport since plasmas often contain multiple ion species. This work focuses on single ion species transport, although has been extended to ionic mixtures. Ion transport becomes magnetized when the ion Hall parameter $\chi_i = \omega_{\text{c},i}\tau_{ii} \gtrsim 1$. Since ions are much heavier than electrons, much stronger magnetic fields are required to magnetize ion transport. The local ion heat flux has a similar form to the electron conductive heat flux with new dimensionless coefficients.



The viscous stress tensor is another transport quantity that requires a closure within a hydrodynamic framework. In an ideal plasma, ion viscosity almost always dominates over electron viscosity; therefore, it is considered here in the context of ions. Braginskii's closure for the ion viscous stress tensor, valid for order-unity Mach number, can be expressed in the form 
\begin{align}
    \Pi_i^\text{local} &= - \eta_c^i \Big( \underbrace{\hat{\eta}_0^i \mathrm{W}_0^i+ \hat{\eta}_1^i \mathrm{W}_1^i  +  \hat{\eta}_2^i \mathrm{W}_2^i}_{\text{Ordinary viscous stress}} \underbrace{ -\hat{\eta}_3^i \mathrm{W}_3^i  - \hat{\eta}_4^i \mathrm{W}_4^i}_{\text{Gyroviscous stress}} \Big)
\end{align}
where $\eta_c^i = n_i T_i \tau_{ii}$ is the basic viscosity and $\hat{\eta}^i_{(0,\dots,4)}$ are a set of dimensionless viscosity coefficients. $\mathrm{W}_{(0,\dots,4)}^i$ are various contractions of $\boldsymbol{b}$ with the ion rate-of-strain tensor $\mathrm{W}^i = \boldsymbol{\nabla}\boldsymbol{u}_i + (\boldsymbol{\nabla}\boldsymbol{u}_i)^{\text{T}} - \frac{2}{3}(\boldsymbol{\nabla}\cdot\boldsymbol{u}_i)\mathbb{I}$, defined by Braginskii as
\begin{align}
    \mathrm{W}_{0,ab}^i &= \frac{3}{2}\bigg(b_a b_b - \frac{1}{3}\delta_{ab}\bigg)
        \bigg(b_c b_d - \frac{1}{3}\delta_{cd}\bigg)\mathrm{W}^i_{cd}, \nonumber\\
    \mathrm{W}_{1,ab}^i &= \bigg( \delta_{ac}^\perp \delta_{bd}^\perp
        + \frac{1}{2}\delta_{ab}^\perp b_c b_d \bigg)\mathrm{W}^i_{cd}, \nonumber\\
    \mathrm{W}_{2,ab}^i &= \bigg( \delta_{ac}^\perp b_b b_d
        + \delta_{bd}^\perp b_a b_c \bigg)\mathrm{W}^i_{cd}, \nonumber\\
    \mathrm{W}_{3,ab}^i &= \frac{1}{2}\bigg( \delta_{ac}^\perp \varepsilon_{bed}
        + \delta^\perp_{bd}\varepsilon_{aec} \bigg)h_e \mathrm{W}^i_{cd}, \nonumber\\
    \mathrm{W}_{4,ab}^i &= \bigg( b_a b_c \varepsilon_{bed}
        + b_b b_d \varepsilon_{aec} \bigg)h_e \mathrm{W}^i_{cd}. 
\end{align}
where index summation convention is employed, $\delta^\perp_{ab} \equiv \delta_{ab}-b_a b_b$ is the rejection dyad, and $\varepsilon_{abc}$ is the Levi-Civita symbol. Similarly to the vectorial transport fluxes, this local closure contains parallel, perpendicular, and cross terms \cite{stacey_viscous_1985}. The coefficient $\hat{\eta}_0^i$ is independent of the Hall parameter since it represents the parallel components unaffected by the magnetic field. The $\hat{\eta}_{(1,2)}^i$ terms represent the perpendicular components of the viscous stress tensor suppressed by the magnetic field. The $\hat{\eta}_{(3,4)}^i$ terms are the additional cross components gained due to gyromotion, analogously to the cross components of vectorial transport fluxes such as the Righi-Leduc term.

\section{Reduced Kinetic Method}\label{section_model}

\subsection{Governing equations}

The \ac{VFP} equation provides a kinetic description of an ideal weakly coupled plasma, accounting for long-range Coulomb collisions through a Fokker-Planck collision operator derived from first principles. The electron \ac{VFP} equation is
\begin{equation}\label{electron_VFP}
\begin{aligned}
    \partial_t f_e &+ \boldsymbol{v}\cdot \boldsymbol{\nabla}_{\boldsymbol{x}} f_e - \frac{e}{m_e}\big(\boldsymbol{E} + \boldsymbol{v}\times\boldsymbol{B}\big)\cdot\boldsymbol{\nabla}_{\boldsymbol{v}}f_e \\&= C_{ee}\{f_e,f_e\} + C_{ei}\{f_e,f_i\},
\end{aligned}
\end{equation}
where $(t,\boldsymbol{x},\boldsymbol{v})$ are the laboratory frame phase space coordinates for time, particle position, and particle velocity, respectively. $f_e(t,\boldsymbol{x},\boldsymbol{v})$ is the electron \ac{DF}, and $\boldsymbol{E}$ and $\boldsymbol{B}$ are the macroscopic electric and magnetic fields. The terms $ C_{ee}\{f_e,f_e\} $ and $C_{ei}\{f_e,f_i\}$ are the Fokker-Planck collision operators for electron-electron and electron-ion collisions respectively.

The derivation of the \ac{RKM} proceeds similarly to previous work \cite{Mitchell_2024}: the electron \ac{VFP} equation is transformed into the electron \ac{CoM} frame by transforming the velocity coordinate to the peculiar velocity $\boldsymbol{w}=\boldsymbol{v}-\boldsymbol{u}_e$. The electron \ac{DF} is then decomposed as $f_e = f^M_e + \delta f_e$ to find the kinetic equation for the \ac{DDF}, with Maxwellian terms expressed in terms of thermodynamic forces. The nonlinear $C_{ee}\{\delta f_e,\delta f_e\}$ component of the collision operator is neglected with this assumption checked \textit{a posteriori}, leaving a linear inhomogeneous equation for $\delta f_e$. Finally, the \ac{DDF} is expanded over spherical harmonics in velocity space and considered in slab planar spatial geometry, leaving a set of coupled PDEs for the modes $\delta f_e^{(l,m)}(z,w)$. Solving these PDEs for given hydrodynamic profiles, the \ac{DDF} is recovered, from which transport fluxes may be evaluated.

\subsection{Derivation of DDF equation}

Starting with the electron VFP equation \ref{electron_VFP}, we change velocity coordinates to the frame comoving with the local flow, i.e. $(t,\boldsymbol{x},\boldsymbol{v})$ to $(t'=t,\boldsymbol{x}'=\boldsymbol{x},\boldsymbol{w}=\boldsymbol{v}-\boldsymbol{u}_e(t,\boldsymbol{x}) )$. Derivatives transform as
\begin{equation}
\begin{aligned}
 \partial_t &= \partial_{t'} - (\partial_t \boldsymbol{u}_e)\cdot\boldsymbol{\nabla}_{\boldsymbol{w}},\\
 \boldsymbol{\nabla}_{\boldsymbol{x}} = \boldsymbol{\nabla}_{\boldsymbol{x}'} - &(\boldsymbol{\nabla}_{\boldsymbol{x}} \boldsymbol{u}_e)\cdot\boldsymbol{\nabla}_{\boldsymbol{w}}
,\quad \boldsymbol{\nabla}_{\boldsymbol{v}} = \boldsymbol{\nabla}_{\boldsymbol{w}}.
\end{aligned}
\end{equation}
Dropping primes, this yields the electron VFP equation in the electron \ac{CoM} frame as
\begin{equation}\label{kinetic_equation_com}
\begin{aligned}
    &\partial_t f_e + (\boldsymbol{w}+\boldsymbol{u}_e)\cdot\boldsymbol{\nabla}f_e - \bigg( \partial_t \boldsymbol{u}_e + (\boldsymbol{w}+\boldsymbol{u}_e)\cdot (\boldsymbol{\nabla}\boldsymbol{u}_e)\\& + \frac{e}{m_e}\big( \boldsymbol{E}+ (\boldsymbol{w}+\boldsymbol{u}_e)\times\boldsymbol{B}\big) \bigg)\cdot\boldsymbol{\nabla}_{\boldsymbol{w}} f_e 
    \\&=C_{ee}\{f_e,f_e\} + C_{ei}\{f_e,f_i\}.
\end{aligned}
\end{equation}

The electron momentum equation, obtained as the $m_e \boldsymbol{v}$ moment of the electron kinetic equation \ref{electron_VFP} or $m_e\boldsymbol{w}$ moment of Equation \ref{kinetic_equation_com}, is
\begin{equation}\label{electron_flow_equation}
\begin{aligned}    
    &\partial_t \boldsymbol{u}_e + \boldsymbol{u}_e\cdot\boldsymbol{\nabla}\boldsymbol{u}_e \\&= \frac{1}{\rho_e} \bigg( -\boldsymbol{\nabla} p_e -\boldsymbol{\nabla}\cdot \Pi_e - en_e \big( \boldsymbol{E} + \boldsymbol{u}_e \times \boldsymbol{B} \big) + \boldsymbol{R}_e \bigg),
\end{aligned}
\end{equation}
where $p_e = n_e T_e$ is the electron thermal pressure, $\Pi_e$ is the electron viscous stress tensor, and $\boldsymbol{R}_e$ is the electron-ion momentum exchange defined as $\boldsymbol{R}_e = \int d^3\boldsymbol{v}\ m_e \boldsymbol{v} C_{ei}\{f_e\}$. Substituting in $\partial_t \boldsymbol{u}_e$ from the electron momentum equation \ref{electron_flow_equation} into Equation \ref{kinetic_equation_com} yields

\begin{equation}\label{kinetic_equation_com2}
    \partial_t f_e + \hat{V}_e f_e + \hat{B}_e f_e = C_{ee}\{f_e,f_e\} + C_{ei}\{f_e,f_i\},
\end{equation}
where the unmagnetized electron Vlasov operator in the \ac{CoM} frame
\begin{equation}
\begin{aligned}    
    &\hat{V}_e = (\boldsymbol{w}+\boldsymbol{u}_e)\cdot \boldsymbol{\nabla} \\&- \bigg[ \boldsymbol{w}\cdot  (\boldsymbol{\nabla} \boldsymbol{u}_e)  + \frac{1}{\rho_e}\bigg( -\boldsymbol{\nabla} p_e -\boldsymbol{\nabla}\cdot \Pi_e  + \boldsymbol{R}_e\bigg) \bigg] \cdot\boldsymbol{\nabla}_{\boldsymbol{w}},
\end{aligned}
\end{equation}
and the magnetic field operator
\begin{equation}
    \hat{B}_e = -  \frac{e}{m_e}(\boldsymbol{w}\times\boldsymbol{B})\cdot\boldsymbol{\nabla}_{\boldsymbol{w}}.
\end{equation}
The electron \ac{DF} is decomposed as $f_e = f^M_e + \delta f_e$, where the electron Maxwellian distribution is $f^M_e = n_e (\sqrt{\pi} v_{\text{th},e})^{-3} \exp( - (w/v_{\text{th},e})^2) $ with thermal velocity $v_{\text{th},e} = \sqrt{2T_e/m_e}$. Unlike the Chapman-Enskog expansion, this decomposition does not assume a small deviation from Maxwellian $\vert \delta f_e / f^M_e \vert \ll 1$. This expands the collision terms as

\begin{equation}
\begin{aligned}
    C_{ee}\{ f_e , f_e\} &\approx \underbrace{C_{ee}\{ \delta f_e,f^M_e\}}_{=\hat{C}_{ee}^T\delta f_e} + \underbrace{C_{ee}\{ f^M_e , \delta f_e \}}_{=\hat{C}_{ee}^F\delta f_e}, 
    \\ C_{ei}\{ f_e , f_i\} &\approx \underbrace{C_{ei} \{ f^M_e , f^M_i \}}_{=C^{M}_{ei}} + \underbrace{C_{ei}\{ \delta f_e , f^M_i\}}_{ = \hat{C}_{ei}\delta f_e},
\end{aligned}
\end{equation}
where $C_{ee}\{f^M_e,f^M_e\} = 0$ identically and the nonlinear electron-electron term $C_{ee}\{\delta f_e,\delta f_e\}$ may be neglected even for order-unity deviations from Maxwellian in the tail of the distribution. This assumption is checked \textit{a posteriori} in Section \ref{a_posteriori_check}. For velocities relevant to electron transport $w \sim v_{\text{th},e} \gg v_{\text{th},i}$, the electron-ion collision operator to leading order in mass ratio $m_e / m_i \ll 1$ reduces to the same form as colliding against a Maxwellian ion distribution, i.e. $C_{ei}\{f_e , f_i\} \approx C_{ei}\{f_e,f^M_i\}$. Even if multiple ion species are present or the plasma is not quasineutral, the electron-ion collision operator against individual ion species is in a form that allows the total electron-ion collision operator to be written in the same form as a single-species electron-ion collision operator with effective ionization $Z = \sum_{\beta} \hat{\nu}_{e\beta} / \hat{\nu}_{ee} = \sum_{\beta} Z_{\beta}^2 \frac{n_{\beta}}{n_e}  \frac{\ln\Lambda_{e\beta}}{\ln\Lambda_{ee}}  $ for ions with labels $\beta$. Equation \ref{kinetic_equation_com2} becomes
\begin{equation}\label{kinetic_equation_com3}
    \partial_t f^M_e + \hat{V}_e f^M_e -  C_{ei}^M = \big( \hat{C}_{e}- \partial_t - \hat{V}_e - \hat{B}_e \big)\delta f_e,
\end{equation}
where $\hat{C}_e = \hat{C}_{ee}^T + \hat{C}_{ee}^F  + \hat{C}_{ei}$.

On hydrodynamic timescales, transport fluxes are typically quasistatic, i.e. are only dependent on time implicitly through hydrodynamic profiles \cite{10.1063/5.0134966}. Since transport quantities are evaluated from $\delta f_e$, we take $\partial_t \delta f_e = 0$ in this work. In a fluid simulation, the effect of the history of the \ac{DDF} may still be taken into account by writing 
\begin{equation}
    \partial_t \delta f_e \approx \frac{\delta f_e(t) - \delta f_e (t - \Delta t)}{\Delta t},
\end{equation}
where $\Delta t$ is the current hydrodynamic timestep, and $\delta f_e (t-\Delta t)$ has already been determined by the \ac{RKM} on the previous timestep. The $\delta f_e(t-\Delta t) / \Delta t$ term is then moved to the left hand side into the driving terms, and $\delta f_e(t) / \Delta t$ is kept on the right hand side as an operator. For the purposes of this paper, these terms are ignored, but could be straightforwardly included when integrating the \ac{RKM} into a fluid simulation.

The Maxwellian terms on the left hand side may be evaluated using the time derivative
\begin{equation}
\begin{aligned}
    &\partial_t f^M_e = \bigg( \partial_t \ln n_e + \bigg( \frac{w^2}{v_{\text{th},e}^2} - \frac{3}{2} \bigg) \partial_t \ln T_e \bigg) f^M_e
    \\ &= \bigg[ -  \boldsymbol{\nabla} \cdot \boldsymbol{u}_e - \boldsymbol{u}_e \cdot \boldsymbol{\nabla}  \ln n_e + \bigg( \frac{w^2}{v_{\text{th},e}^2} - \frac{3}{2} \bigg)\\&\times \bigg( -\boldsymbol{u}_e\cdot\boldsymbol{\nabla}  \ln T_e - \frac{2}{3}\boldsymbol{\nabla} \cdot\boldsymbol{u}_e + K_e \bigg) \bigg] f^M_e ,
\end{aligned}
\end{equation}
and gradients
\begin{equation}
\begin{aligned}
    &\boldsymbol{\nabla}  f^M_e = \bigg[ \boldsymbol{\nabla}  \ln n_e + \bigg( \frac{w^2}{v_{\text{th},e}^2} - \frac{3}{2} \bigg)\boldsymbol{\nabla}  \ln T_e \bigg] f^M_e,
    \\ &\boldsymbol{\nabla}_{\boldsymbol{w}} f^M_e = - \frac{2}{v_{\text{th},e}^2} f^M_e \boldsymbol{w}.
\end{aligned}
\end{equation}
The time derivatives of the density and temperature have been rewritten using the electron density and temperature fluid equations which may be written in the form
\begin{equation}
\begin{aligned}
    \partial_t \ln n_e &= -  \boldsymbol{\nabla} \cdot \boldsymbol{u}_e - \boldsymbol{u}_e \cdot \boldsymbol{\nabla}  \ln n_e ,
    \\ \partial_t \ln T_e &= - \boldsymbol{u}_e\cdot\boldsymbol{\nabla}  \ln T_e - \frac{2}{3}\boldsymbol{\nabla} \cdot\boldsymbol{u}_e +  K_e,
\end{aligned}
\end{equation}
with the transport terms are written together as
\begin{equation}
\begin{aligned}
    K_e &= \frac{2}{3 p_e}\bigg[ -\boldsymbol{\nabla} \cdot\boldsymbol{q}_e   - \Pi_e\colon \boldsymbol{\nabla} \boldsymbol{u}_e - \boldsymbol{R}_e\cdot \Delta \boldsymbol{u}
    \bigg]. 
\end{aligned}
\end{equation}
Combining these with the Maxwellian-Maxwellian collision term
\begin{equation}
    C_{ei}^M= -2\hat{\nu}_{ei} v_{\text{th},e} \frac{\Delta\boldsymbol{u}\cdot\boldsymbol{w}}{w^3} f^M_e,
\end{equation}
the inhomogeneous driving terms on the left hand side of Equation \ref{kinetic_equation_com3} become
\begin{equation}
\begin{aligned}
    D_e &= \bigg[ \bigg( \frac{w^2}{v_{\text{th},e}^2} - \frac{3}{2} \bigg)  K_e   + \frac{2}{v_{\text{th},e}^2}  \boldsymbol{\nabla}\boldsymbol{u}_e \colon \bigg( \boldsymbol{w}\boldsymbol{w} - \frac{1}{3}w^2\mathbb{I} \bigg) \\& +   \bigg( \bigg( \frac{w^2}{v_{\text{th},e}^2} - \frac{5}{2}  \bigg)  \boldsymbol{\nabla} \ln T_e + \frac{1}{p_e}\bigg( -\boldsymbol{\nabla}\cdot \Pi_e  + \boldsymbol{R}_e\bigg) \\&+ 2\hat{\nu}_{ei}  \frac{v_{\text{th},e}}{w^3}\Delta\boldsymbol{u}\bigg) \cdot \boldsymbol{w} \bigg] f^M_e.
\end{aligned}
\end{equation}
The electron \ac{VFP} equation in the electron frame for the \ac{DDF} $\delta f_e$ is then
\begin{equation}\label{electron_DDF_equation}
    D_e  = \big( \hat{C}_{e} - \hat{V}_e - \hat{B}_e \big)\delta f_e.
\end{equation}
with the implemented form of the linear collision operators given in Section \ref{linear_collision_operators}. The \ac{DDF} equation in the \ac{CoM} frame is solved with the same solvability conditions as the Chapman-Enskog solution
\begin{equation}\label{solvability_conditions}
    \int d^3\boldsymbol{w}\ \{1 , \boldsymbol{w}, w^2\}\, \delta f_e = 0,
\end{equation}
such that the solution does not modify the input density, flow, or temperature. Once Equation \ref{electron_DDF_equation} is solved for the \ac{DDF} $\delta f_e$, transport quantities may be computed as
\begin{equation}
\begin{aligned}
    \boldsymbol{q}_\alpha &= \int d^3\boldsymbol{w}\ \frac{1}{2}m_\alpha w^2\boldsymbol{w} \delta f_\alpha,
    \\\boldsymbol{R}_e &= \boldsymbol{R}_e^0 + \int d^3\boldsymbol{w}\ m_e\boldsymbol{w}\hat{C}_{ei}\delta f_e,
    \\\Pi_\alpha &= \int d^3\boldsymbol{w}\ m_\alpha\bigg( \boldsymbol{w}\boldsymbol{w}-\frac{1}{3}w^2 \mathbb{I}\bigg) \delta f_\alpha,
\end{aligned}
\end{equation}
where $\boldsymbol{R}_e^0 = -\frac{3}{4\sqrt{\pi}}\rho_e \hat{\nu}_{ei} \Delta \boldsymbol{u} $ is the dynamic friction between an electron and ion Maxwellian. When the Vlasov term $\hat{V}_e \delta f_e$ is neglected, the \ac{DDF} Equation \ref{electron_DDF_equation} reduces to the leading-order Chapman-Enskog equation for $\delta f_e$, therefore naturally reproduces local analytic transport results for small Knudsen number.

\subsection{Form of collision operator}\label{linear_collision_operators}

The Fokker-Planck collision operator is written in Rosenbluth form as
\begin{equation}\begin{aligned}
    C_{\alpha\beta}\{ f_\alpha,f_\beta\} &=L_{\alpha\beta} \frac{\partial}{\partial v_i}\bigg[ -\frac{m_\alpha}{m_\beta} \frac{\partial \mathcal{H}_\beta}{\partial v_i}  f_\alpha + \frac{1}{2} \frac{\partial^2 \mathcal{G}_\beta}{\partial v_i \partial v_j} \frac{\partial f_\alpha }{\partial v_j}  \bigg],
\end{aligned}\end{equation}
where $L_{\alpha\beta}=\frac{1}{4\pi}\big(\frac{q_\alpha q_\beta}{\varepsilon_0 m_\alpha}\big)^2\ln\Lambda_{\alpha\beta}$ and the Rosenbluth potentials are defined as
\begin{equation}\begin{aligned}\label{eqn:rosenbluth_potentials}
   \mathcal{H}_\beta \equiv \int d^3\boldsymbol{v}'\ \frac{1}{u} f_\beta(\boldsymbol{v}'), \quad \mathcal{G}_\beta \equiv \int d^3\boldsymbol{v}'\ u f_\beta(\boldsymbol{v}')
\end{aligned}\end{equation}
with $u=\vert \boldsymbol{v}-\boldsymbol{v}'\vert$. The test-particle and field-particle collision operators are then defined as $\hat{C}^T_{\alpha\beta} \delta f_\alpha = C_{\alpha\beta}\{\delta f_\alpha, f^M_\beta\}$ and $\hat{C}^F_{\alpha\beta}\delta f_\beta = C_{\alpha\beta}\{f_\alpha^M,\delta f_\beta\}$ respectively.

The test-particle collision operator, which appears for both like-like and like-unlike collisions, has the form
\begin{equation}
    \label{C_alpha_beta}
\begin{aligned}    \hat{C}^T_{\alpha\beta}\delta f_\alpha &= \frac{1}{2} \nu_D^{\alpha\beta} \hat{\mathcal{L}}\delta f_\alpha \\&+ \frac{1}{w^2} \partial_w \bigg( \tilde{\nu}_s^{\alpha\beta} w^3\delta f_\alpha + \frac{1}{2}\nu_\parallel^{\alpha\beta} w^4 \partial_w\delta f_\alpha \bigg),
\end{aligned}
\end{equation}
where the three collision frequencies are defined via
\begin{equation}\label{collision_frequencies}
    \begin{aligned}
         \nu^{\alpha\beta}_D &= \hat{\nu}_{\alpha\beta}\frac{\text{erf}(\xi_\beta) - G(\xi_\beta)}{\xi_\alpha^3},
         \\ \tilde{\nu}_s^{\alpha\beta} &= \frac{m_\alpha}{m_\alpha + m_\beta} \nu^{\alpha\beta}_s = 2\hat{\nu}_{\alpha\beta}\frac{G(\xi_\beta)}{\xi_\alpha},
        \\ \nu_\parallel^{\alpha\beta} &= 2\hat{\nu}_{\alpha\beta}\frac{G(\xi_\beta)}{\xi_\alpha^3}
        \,,
    \end{aligned}
\end{equation}
where $\xi_\alpha = w / v_{\text{th},\alpha}$, $G(\xi)$ is the Chandrasekhar function \cite{chandrasekhar_dynamical_1943,helander_and_sigmar}, and the pitch-angle scattering operator
\begin{equation}
    \hat{\mathcal{L}} f = (\boldsymbol{w}\times\boldsymbol{\nabla}_{\boldsymbol{w}})^2 f=\partial_\mu \big( (1-\mu^2) \partial_\mu f \big) + \frac{1}{1-\mu^2}\partial_\varphi^2 f .
\end{equation}
The field-particle collision operator has the form
\begin{equation} \begin{aligned}
    \hat{C}^F_{\alpha\beta} \delta f_\beta =  &\bigg(\frac{q_\alpha q_\beta}{\varepsilon_0 m_\alpha} \bigg)^2 \ln \Lambda_{\alpha\beta}\frac{m_\alpha f^M_\alpha}{T_\alpha}  \Biggl[ \frac{ T_\alpha}{m_\beta} \delta f_\beta  -\delta  \mathcal{H}_\beta \\&
    + \bigg(\frac{m_\alpha}{m_\beta} - 1 \bigg) w \partial_w\delta  \mathcal{H}_\beta + \frac{m_\alpha w^2}{2 T_\alpha} \partial_w^2\delta  \mathcal{G}_\beta \Biggl],
\end{aligned}\end{equation} 
where $\delta \mathcal{H}_\beta$ and $\delta \mathcal{G}_\beta$ are the Rosenbluth potentials evaluated from $\delta f_\beta$. 

The electron-ion collision operators have the form 
\begin{equation}
\begin{aligned}
    \hat{C}_{ei} \delta f_e &= \hat{\nu}_{ei} \frac{1}{2\xi_e^3} \hat{\mathcal{L}}  \delta f_e,
    \quad  C_{ei}^M= -2\hat{\nu}_{ei} v_{\text{th},e} \frac{\Delta\boldsymbol{u}\cdot\boldsymbol{w}}{w^3} f^M_e,
\end{aligned}
\end{equation}
where $\Delta \boldsymbol{u} = \boldsymbol{u}_e - \boldsymbol{u}_i$. For the \ac{RKM} for a single ion species, the ion-electron collision operators \cite{Braginskii1965ReviewsOP} have the form
\begin{equation}
\begin{aligned}
    C_{ie}\{\delta f_i,f_e\} &= \frac{1}{\rho_i}\boldsymbol{R}_e\cdot\boldsymbol{\nabla}_{\boldsymbol{w}}\delta f_i + \hat{C}^T_{ie}\delta f_i, 
    \\ C_{ie}^M &= \bigg[ - \frac{1}{p_i} \boldsymbol{R}_e \cdot\boldsymbol{w}\\& - \frac{8}{3\sqrt{\pi}} \hat{\nu}_{ei} \frac{\rho_e}{\rho_i} \bigg(1-\frac{T_e}{T_i}\bigg) \bigg(  \frac{w^2}{v_{\text{th},i}^2}- \frac{3}{2}  \bigg) \bigg]f^M_i,
\end{aligned}
\end{equation}
where $\boldsymbol{w}=\boldsymbol{v}-\boldsymbol{u}_i$ for the ion equations.

\subsection{Spherical harmonic expansion}

The \ac{DDF} is then expanded in spherical harmonics in velocity space as

\begin{equation}\label{spherical_harmonic_expansion}
    \delta f_e= \sum_{k=0}^\infty \sum_{n=-k}^k \frac{2k+1}{4\pi}\frac{(k-n)!}{(k+n)!}P_k^n(\cos\vartheta)e^{in\varphi} \delta f_e^{(k,n)},
\end{equation}
where $P_k^n$ are the associated Legendre functions \cite{noauthor_dlmf_nodate}, and $\vartheta,\varphi$ are the polar and azimuthal velocity coordinates respectively. The orientation of the spherical velocity coordinates is chosen such that the pole ($\vartheta=0$) is aligned with the $z$ axis via $\mu =\cos\vartheta = \boldsymbol{w}\cdot\boldsymbol{\hat{z}}/ w$, and $\varphi =0$ is contained within the $xz$ plane via $\cos\varphi = \boldsymbol{w}\cdot\boldsymbol{\hat{x}}/ w\sin\vartheta$. The components of the spherical harmonic expansion may be recovered by an appropriate moment of the \ac{DDF} over velocity angle space

\begin{equation}
    \delta f_e^{(l,m)} = \int d^2\Omega\ e^{-im\varphi}P_l^m(\mu) \delta f_e = \langle l,m\vert  \delta f_e\rangle ,
\end{equation}
for $l,m \in \mathbb{Z}$ with $l \geq 0$ and $\vert m \vert \leq l$, where $d^2\Omega = \sin\vartheta d\vartheta d\varphi = -d\mu d\varphi$ is the velocity angle element and Dirac notation is introduced for brevity. The \ac{DDF} equation may be projected onto the spherical harmonic basis by taking the same moment $\int d^2\Omega\  e^{-in\varphi}P_k^n(\mu) (\dots) $, yielding a set of PDEs for the modes of the \ac{DDF} of the form
\begin{equation}\label{electron_DDF_SH_eqn}
    \mathcal{D}_e^{(l,m)}  = \mathcal{C}_{e}^{(l,m)}  - \mathcal{V}_{e}^{(l,m)} - \mathcal{B}_e^{(l,m)},
\end{equation}
where calligraphic characters represent the projections of the corresponding terms onto the spherical harmonic basis. The form of most of these projections are provided by Swanekamp et al. \cite{10.1063/1.5109430}, with other terms provided in \ref{appendix_projection}. The spherical harmonic expansion is then truncated to the first $N_l$ modes for the $l$ index, leaving $N_l^2$ PDEs in $(z,w)$ that may be solved numerically for $\delta f_e$, and therefore transport quantities. For this work, $N_l = 4$ is used which has been shown in previous work \cite{Mitchell_2024,McDevitt_2} to demonstrate good convergence of transport quantities for Knudsen numbers $N_K \lesssim 1/10$ and has been checked again with convergence studies. The solvability conditions (Equation \ref{solvability_conditions}) become
\begin{equation}\begin{aligned}
    \int_0^\infty dw\ w^2 \delta f_e^{(0,0)} &= 0, \quad \int_0^\infty dw\ w^4 \delta f_e^{(0,0)} = 0,
    \\\int_0^\infty dw\ w^3 \delta f_e^{(1,m)} &= 0, \quad \text{for }\ m=-1,0,1,
\end{aligned}\end{equation}

In principle, the DDF equation expanded in spherical harmonics may be solved in full 3D3V space. Three velocity dimensions are required to fully realize magnetic field effects, however a single spatial dimension is suitable for investigating nonlocal transport. The spatial geometry is therefore reduced to a single spatial dimension $z$, with translational symmetry assumed in the $x$ and $y$ directions, leaving a 1D3V space. 

Once the electron or ion \ac{DDF} equation has been solved for $\delta f_\alpha^{(l,m)}$, with $\alpha = e,i$ respectively, transport terms may be recovered as appropriate velocity moments of $\delta f_\alpha$. Using orthogonality relations for spherical harmonics, these may be expressed directly as one-dimensional integrals of $\delta f_\alpha^{(l,m)}$ over $w$ as
\begin{equation}\label{transport_from_lm}
\begin{aligned}
    \boldsymbol{q}_\alpha &=\frac{1}{2}m_\alpha \int_0^\infty dw\  w^5 \boldsymbol{S}_{1,\alpha},
        \\ \boldsymbol{R}_e &= \boldsymbol{R}_e^0 - \hat{\nu}_{ei} v_{\text{th},e}^3 \int_0^\infty dw\ \boldsymbol{S}_{1,e} ,
    \\\Pi_\alpha &=m_\alpha\int_0^\infty dw\  w^4  \mathrm{S}_{2,\alpha},
\end{aligned}
\end{equation}
where
\begin{equation}
\begin{aligned}
&\boldsymbol{S}_{1,\alpha} = -\text{Re}\{ \delta f_\alpha^{{(1,1)}} \} \boldsymbol{\hat{x}} + \text{Im}\{ \delta f_\alpha^{{(1,1)}} \} \boldsymbol{\hat{y}} + \delta f_\alpha^{{(1,0)}}  \boldsymbol{\hat{z}},\\
&\mathrm{S}_{2,\alpha} = \bigg(
        \frac{1}{12} \text{Re}\{\delta f_\alpha^{(2,2)}\} - \frac{1}{3}\delta f_\alpha^{(2,0)} \bigg) \boldsymbol{\hat{x}}\boldsymbol{\hat{x}}
        \\& + \bigg( -
        \frac{1}{12} \text{Re}\{\delta f_\alpha^{(2,2)}\} - \frac{1}{3}\delta f_\alpha^{(2,0)} \bigg) \boldsymbol{\hat{y}}\boldsymbol{\hat{y}}
        \\& +\frac{2}{3} \delta f_\alpha^{(2,0)}\boldsymbol{\hat{z}}\boldsymbol{\hat{z}}
         -\frac{1}{12} \text{Im}\{\delta f_\alpha^{(2,2)}\} ( \boldsymbol{\hat{x}}\boldsymbol{\hat{y}} + \boldsymbol{\hat{y}}\boldsymbol{\hat{x}}   ) 
        \\&  - \frac{1}{6}\text{Re}\{\delta f_\alpha^{(2,1)}\}( \boldsymbol{\hat{x}}\boldsymbol{\hat{z}} + \boldsymbol{\hat{z}}\boldsymbol{\hat{x}}   ) 
         + \frac{1}{6}\text{Im}\{\delta f_\alpha^{(2,1)}\} ( \boldsymbol{\hat{y}}\boldsymbol{\hat{z}} + \boldsymbol{\hat{z}}\boldsymbol{\hat{y}}   )  .
\end{aligned}
\end{equation}

\subsection{Numerical implementation}

Equation \ref{electron_DDF_SH_eqn} is solved numerically on a simple uniform grid in the single spatial coordinate $z= z_{\text{min}} + i\Delta z$ and speed coordinate $w = (j+1)\Delta w$. For spatial derivatives of a quantity $Q$, a simple finite difference scheme $[\partial_z Q]_i \approx (Q_{i+1} - Q_{i-1})/(2\Delta z)$ is used, and for velocity derivatives, a Chang-Cooper-inspired scheme \cite{CHANG_COOPER}. Equation \ref{electron_DDF_SH_eqn} is discretized as $\boldsymbol{D} = \hat{M}\boldsymbol{F}$, where $\boldsymbol{D}$ and $\boldsymbol{F}$ are vectors of the driving term $\mathcal{D}_e^{(l,m)}$ and modes of the \ac{DDF} $\delta f_e^{(l,m)}$ respectively discretized on the uniform grid, and $\hat{M}$ contains the appropriate finite-differenced operators for the spherical harmonic projections of the collision, Vlasov, and magnetic field operators. The discretized equation may then be solved using standard Python libraries for solving linear problems. In contrast with a previous approach to the \ac{RKM}, the \ac{DDF} equation is solved over the entire velocity domain rather than just the tail, and the field-particle operator is included in the matrix $\hat{M}$.

The driving term $D_e$ and \ac{CoM} Vlasov operator $\hat{V}_e$ are dependent on transport quantities which themselves are determined from moments of $\delta f_e$. For Knudsen numbers $N_{\text{K}}  \lesssim 1/10$, almost all of these terms are subleading and may be neglected as they are in the Chapman-Enskog expansion. The exception to this is the momentum exchange $\boldsymbol{R}_e$ appearing in the electron driving term $D_e$, which depends on $\delta f_e$ to leading order. Braginskii \cite{Braginskii1965ReviewsOP} solves this by taking the integral part of $\boldsymbol{R}_e$ acting on $\delta f_e$ to the right hand side, then solving the resulting linear problem. Although a similar approach may be employed here, we keep these transport terms within $D_e$ on the left hand side. In cases with high Knudsen numbers, the appearance of additional transport terms under spatial derivatives appearing in $D_e$ such as $\boldsymbol{\nabla} \cdot \Pi_e$ would introduce much more complicated and dense matrix operators if included in the right hand side, and the appearance of transport quantities in $\hat{V}_e$ yields a nonlinear problem. These complications are only small corrections to the kinetic equations, therefore they may be handled by solving the equations iteratively. The DDF equation is first solved with the local results for transport quantities to obtain the \ac{DDF} $\delta f_e$. From the \ac{DDF}, transport quantities may be recomputed. In the local limit, these are expected to be consistent. In nonlocal cases, the transport quantities may have modifications, so the solution is inconsistent. These computed transport quantities may be then fed back into the same DDF Equation to be solved again. This procedure is iterated until all transport quantities are self-consistent over the spatial domain within some tolerance. In most cases, neglecting $\mathcal{O}(N_K^2)$ terms is sufficient, leaving $D_e \approx D_e^{(0)}$ where
\begin{equation}
\begin{aligned}
   & D_e^{(0)}= \bigg[  \frac{2}{v_{\text{th},e}^2}  \boldsymbol{\nabla}\boldsymbol{u}_e \colon \bigg( \boldsymbol{w}\boldsymbol{w} - \frac{1}{3}w^2\mathbb{I} \bigg) \\& +   \bigg( \bigg( \frac{w^2}{v_{\text{th},e}^2} - \frac{5}{2}  \bigg)  \boldsymbol{\nabla} \ln T_e + \frac{\boldsymbol{R}_e}{p_e}   +2\hat{\nu}_{ei} \frac{ v_{\text{th},e}}{w^3}\Delta \boldsymbol{u}\bigg) \cdot \boldsymbol{w} \bigg] f^M_e.
\end{aligned}
\end{equation}

\section{Electron transport}\label{section_electron_results}

\subsection{Temperature-gradient-driven terms}\label{subsection_temperature_gradient}

To investigate nonlocality of the electron conductive heat flux $\boldsymbol{q}_e^T$ and thermal force $\boldsymbol{R}_e^T$, a simple hydrogen plasma with a temperature gradient of the form
\begin{equation}
    \label{temperature_error_function}
    T_e(z)
    = 
    \frac{T_{\text{left}} + T_{\text{right}}}{2} + \frac{T_{\text{left}} - T_{\text{right}}}{2}
    \text{erf}\left(\dfrac{z}{L}\right)
    \,,
\end{equation}
is considered with $T_{\text{left}} / T_{\text{right}} = 10$ for various Knudsen numbers $N_K = \lambda_0 / L$, where $\lambda_0 = \lambda_{\text{th},e}(z\rightarrow -\infty)$ is the thermal \ac{MFP} at the hot end. An unmagnetized case is considered, then a magnetized case with a uniform magnetic field $\boldsymbol{B}=-B_0 \boldsymbol{\hat{x}}$, parametrized by the electron Hall parameter at the hot end $\chi_0 = \chi_e(z\rightarrow-\infty)$. The electron thermal \ac{MFP} used here is $\lambda_{\text{th},e} = v_{\text{th},e}/ \hat{\nu}_{ee} \sqrt{1+Z} $.

\begin{figure}[htpb!]
    \centering    \includegraphics[width=0.5\textwidth]{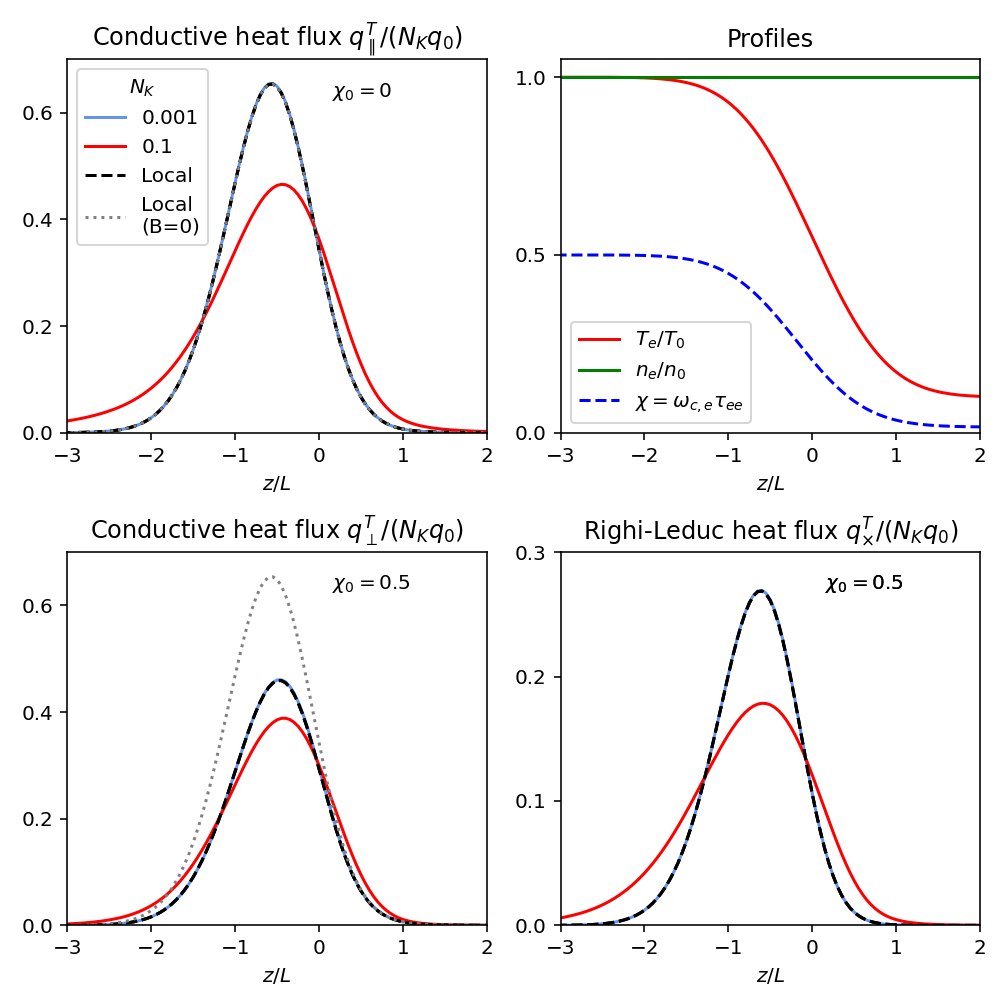}
    \caption{
        Temperature-gradient-driven heat fluxes in an unmagnetized case and moderately magnetized ($\chi_0=0.5$) case, appropriately normalized against Knudsen number and a characteristic heat flux $q_0 = n_0 T_0 v_{0}$, where $n_0$, $T_0$, and $v_0$ are the electron number density, temperature, and thermal velocity at the hot end $z\rightarrow-\infty$, respectively. Local unmagnetized (parallel) results are shown in dotted black lines. The parallel results correspond to the unmagnetized $\chi_0=0$ case. The normalized profiles of temperature, number density, and Hall parameter are shown. The local unmagnetized results in the $z$ direction are shown in black.
    }
    \label{fig:conductive_q}
\end{figure}

Figure \ref{fig:conductive_q} shows the conductive heat flux for a small and large Knudsen number ($N_K=0.001,\, 0.1$ respectively) in an unmagnetized and moderately magnetized case ($\chi_0 = 0,\, 0.5$ respectively). The local analytic result for each component of the conductive heat flux $\boldsymbol{q}_{e,(\parallel,\perp,\times)}^T$ are naturally recovered for small Knudsen number $N_K=0.001$, i.e. the \ac{RKM} solver has no prior knowledge about the local solution, but reproduces it nonetheless for sufficiently small Knudsen number. This is in contrast with many other nonlocal closures which reproduce local transport results by construction of the method. It has been verified that the \ac{RKM} reproduces local analytic results across a wide range of Hall parameters $\chi$ and ionizations $Z$.

For high Knudsen number $N_K = 0.1$, the peak heat flux of each component is reduced for increasing Knudsen number, and a preheat develops in the cold end $0\lesssim z/L \lesssim 1.5$ as expected from physical arguments and \ac{VFP} simulations. For even moderately large Hall parameter $\chi_0 \gtrsim 10$, the magnetic field counteracts nonlocal effects, restoring transport fluxes back to their local analytic results. This is expected physically, since at Hall parameter $\chi_0 \gtrsim 1$, the gyroradius of electrons in the thermal population becomes comparable or smaller than their \acp{MFP}, therefore they may experience more collisions before streaming over sharp temperature gradients.

The Righi-Leduc heat flux $\boldsymbol{q}^T_{e,\times}$ becomes comparable to the other components of the conductive heat flux at intermediate Hall parameter $\chi_0 \sim 1$. At $\chi_0 \lesssim 1$ and high Knudsen number, the Righi-Leduc heat flux is strongly affected by nonlocal effects as has been noted from previous \ac{VFP} simulations \cite{Brodrick_2018}. The Righi-Leduc heat flow reduces thermalization of cold spikes in ICF implosions, therefore has a detrimental impact to the hotspot temperature and fusion yield \cite{PhysRevLett.118.155001}. Nonlocal suppression of Righi-Leduc heat flow may then be beneficial for yield. 

\begin{figure}[htpb!]
    \centering    \includegraphics[width=0.5\textwidth]{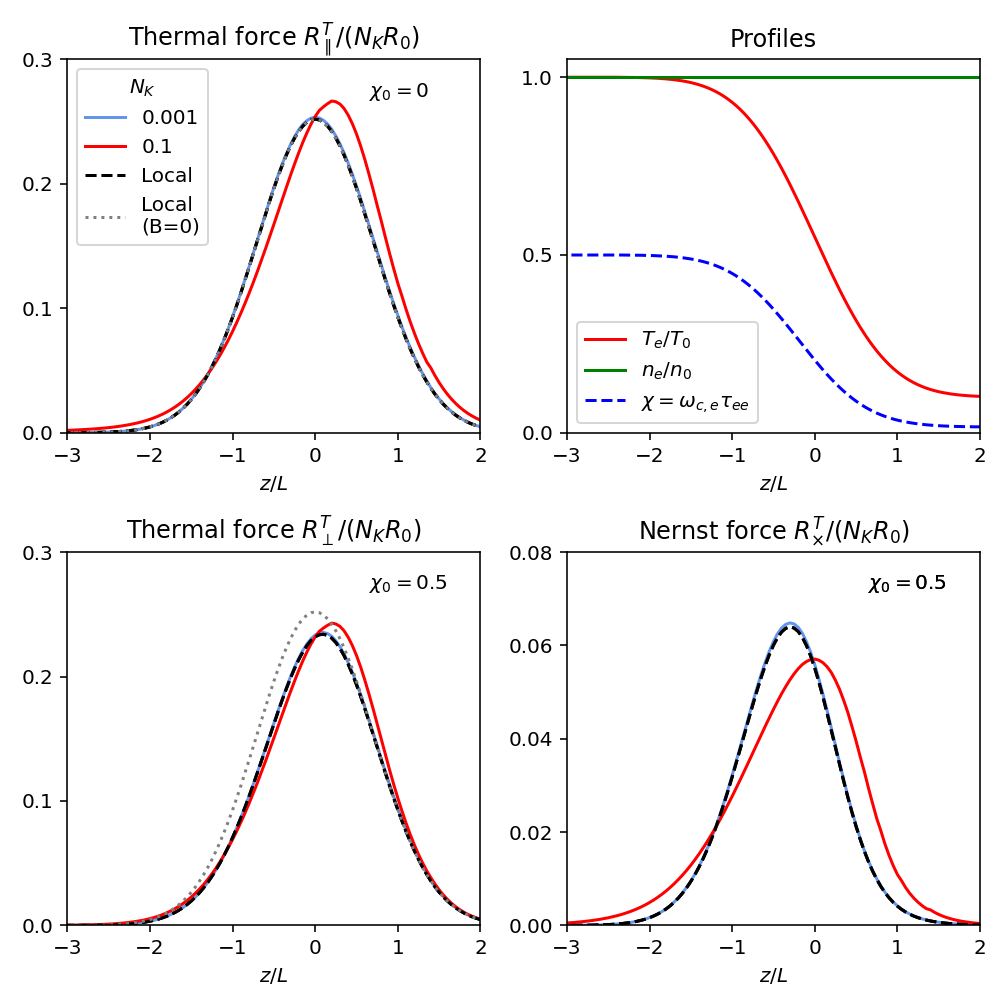}
    \caption{
        Temperature-gradient-driven momentum exchange (thermal force) in an unmagnetized case and moderately magnetized ($\chi_0=0.5$) case, appropriately normalized against Knudsen number and a characteristic momentum exchange $R_0 = n_0 m_e v_0 \hat{\nu}_0$. Local unmagnetized (parallel) results are shown in dotted black lines. The parallel results correspond to the unmagnetized $\chi_0=0$ case. The normalized profiles of temperature, number density, and Hall parameter are shown. The local unmagnetized results in the $z$ direction are shown in black.
    }
    \label{fig:conductive_R}
\end{figure}

Figure \ref{fig:conductive_R} shows the thermal force components $\boldsymbol{R}_{e,(\parallel,\perp,\times)}^T$ for the same cases as Figure \ref{fig:conductive_q}. The thermal force is less impacted by nonlocal effects than the heat flux since it is a lower velocity moment, therefore less sensitive to streaming suprathermal particles. Similarly to the Righi-Leduc heat flow, the Nernst force is significantly more nonlocal compared to its perpendicular counterpart. In ICF implosions, the Nernst effect, originating from $\boldsymbol{R}_e^T$, advects the magnetic field out of the hotspot, therefore demagnetizing thermal transport and having a deleterious impact on maximum temperature and yield. 

Under the standard MHD approximations of neglecting electron inertia and displacement current, the thermal force and friction force enter the generalized Ohm's law as $\boldsymbol{E} = - \frac{1}{en_e} \boldsymbol{R}_e + \dots $, which in turn combines with Faraday's law to form the induction equation $\partial_t \boldsymbol{B} = \boldsymbol{\nabla}\times (\frac{1}{en_e} \boldsymbol{R}_e + \dots) $. Setting $\boldsymbol{B} = 0$ (and therefore $\boldsymbol{j}=0$) to focus on the terms responsible for spontaneously generating magnetic fields, the induction equation becomes $(\partial_t \boldsymbol{B} )\vert_{\boldsymbol{B}=0} = \boldsymbol{\nabla} \times (\frac{1}{en_e}\boldsymbol{R}_{e,\parallel}^T) + \frac{1}{e}\boldsymbol{\nabla} T_e \times \boldsymbol{\nabla}\ln n_e $. The latter term is the Biermann battery, typically viewed as the only MHD term that may spontaneously generate magnetic fields in collisional plasmas. Using the local result for the unmagnetized thermal force, the thermal force term vanishes since $\boldsymbol{\nabla} \times (\frac{1}{e}\hat{\beta}_\parallel \boldsymbol{\nabla} T_e ) \equiv 0$ by vector identity for uniform $Z$. However, the unmagnetized thermal force deviates from this curl-free form due to nonlocal effects, thus this term will not vanish and can spontaneously generate magnetic fields. Nonlocal mechanisms for spontaneous field production have been studied by full \ac{VFP} simulations and other approaches \cite{kingham_nonlocal_2002,ridgers_inadequacy_2021}. This mechanism appears to be related to the generalized Nernst velocity \cite{sherlock_suppression_2020} rather than a nonlocal extension of the Biermann term. This cannot be fully studied in 1D since there must be variation in a second spatial direction, therefore it is not investigated further here.

\subsection{Current-driven terms}

To investigate nonlocality of current-driven transport, we consider an isothermal isochoric simple plasma with a flow in the $y$ direction of the form

\begin{equation}
    \boldsymbol{u}_e = u_0 \exp( - (z/L)^2)\boldsymbol{\hat{y}},
\end{equation}
with a uniform magnetic field in the $-x$ direction $\boldsymbol{B}=-B_0\boldsymbol{\hat{x}}$ and no ion flow $\boldsymbol{u}_i = 0$. The dimensionless number $N_u^0 = \vert u_0 / v_{\text{th},e} \vert $ describes the maximum speed of the flow relative to the electron thermal speed. Previous work has investigated the influence of $N_u^0 \sim 1/100-1/10$ on nonlocal current-driven heat flow \cite{mitchell_nonlocal_2025}, however for simplicity this work focuses on cases with small flows with $N_u^0 = 10^{-4} \ll 1$. The Knudsen number $N_K = \lambda_{\text{th},e} / L$ here defines the width of the flow profile rather than the temperature profile.

\begin{figure}[htpb!]
    \centering    
    \includegraphics[width=0.5\textwidth]{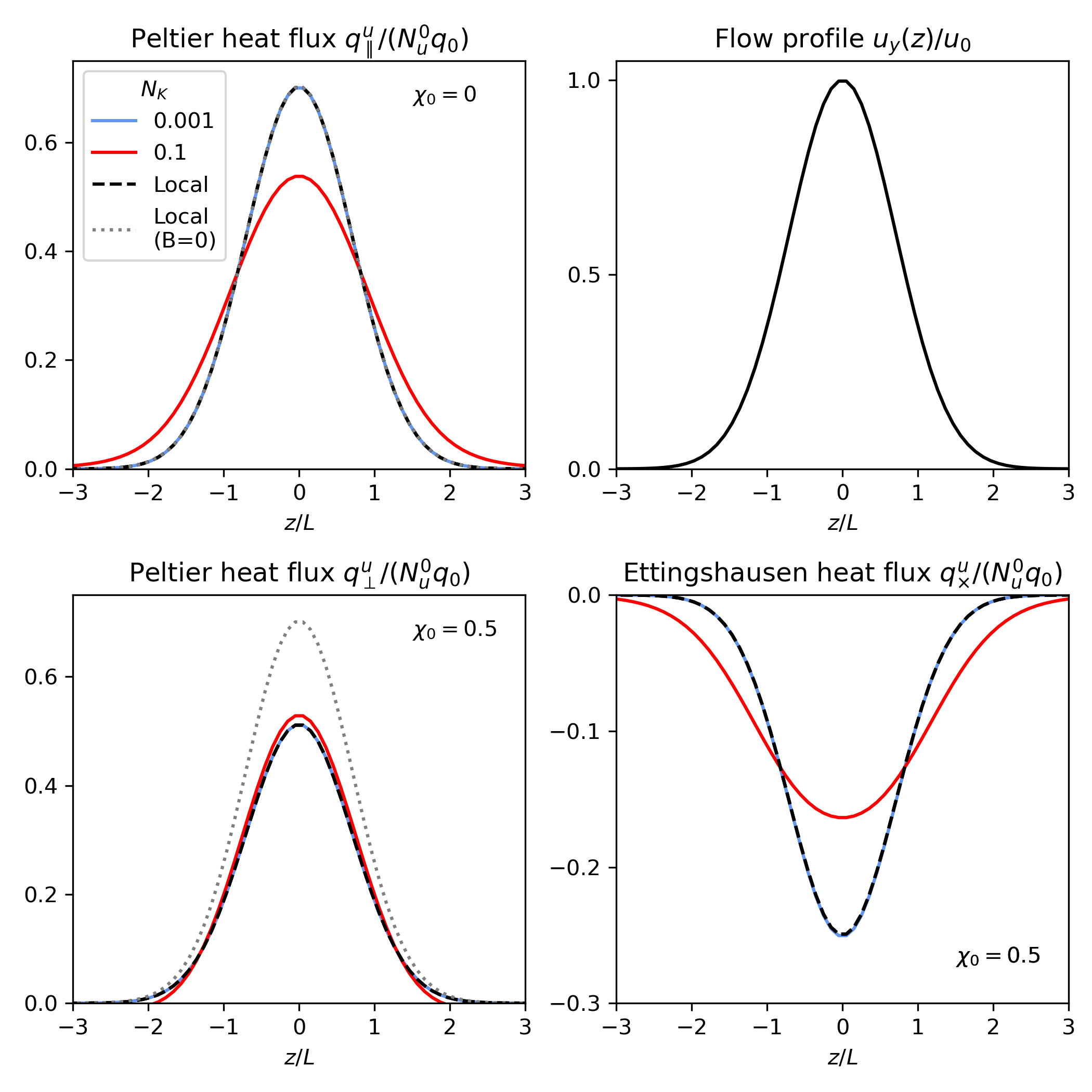}
    \caption{
        Current-driven heat fluxes in a local and nonlocal case for an unmagnetized and moderately magnetized isothermal isochoric simple hydrogen plasma for $N_u^0=10^{-4}\ll1$.
    }
    \label{fig:current_heat_flux}
\end{figure}

\begin{figure}[htpb!]
    \centering    
    \includegraphics[width=0.5\textwidth]{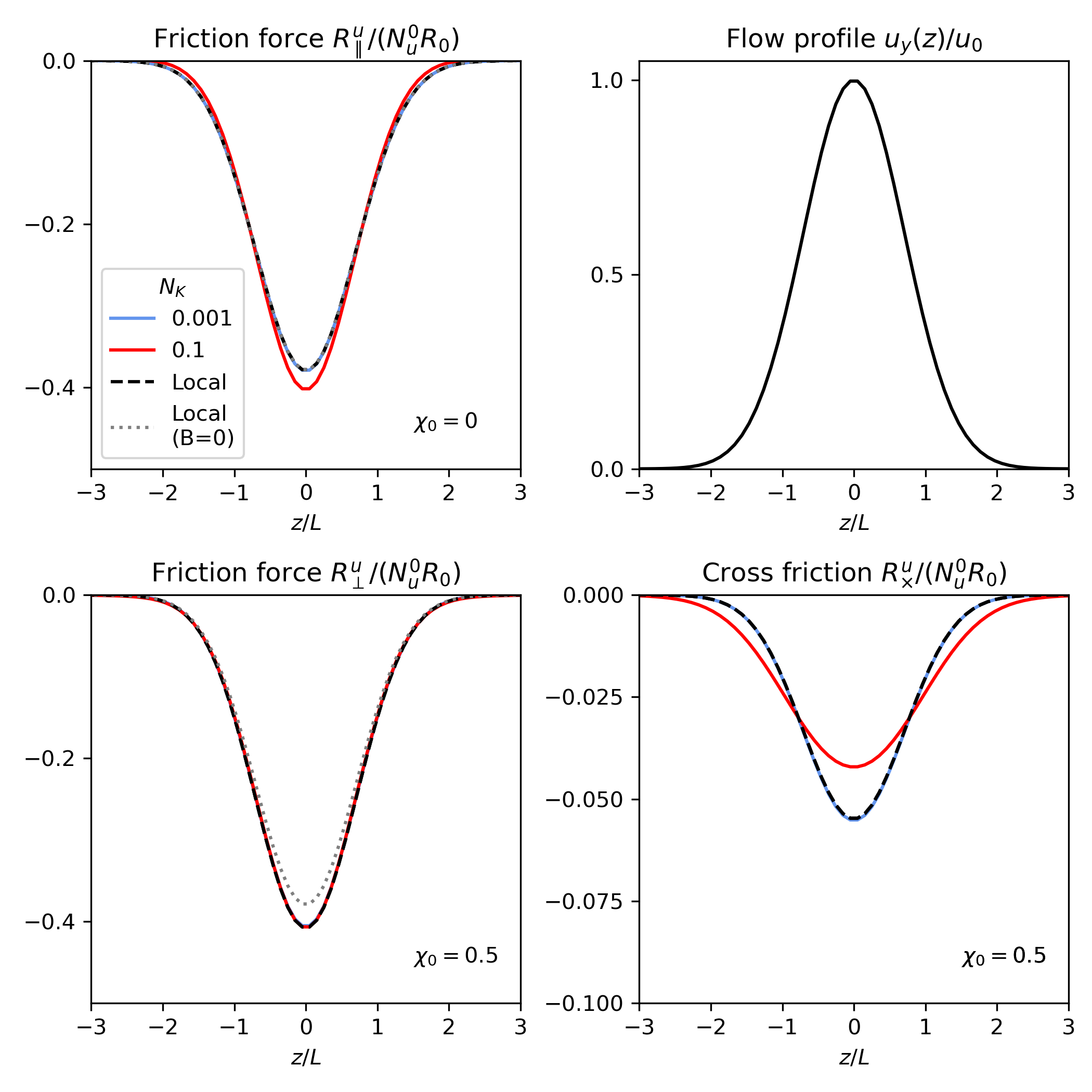}
    \caption{
        Current-driven forces, i.e. friction, in a local and nonlocal case for an unmagnetized and moderately magnetized isothermal isochoric simple hydrogen plasma for $N_u^0=10^{-4}\ll1$.
    }
    \label{fig:current_force}
\end{figure}

The local and nonlocal current-driven heat fluxes (Peltier and Ettingshausen) and forces (friction and cross friction) are shown in Figures \ref{fig:current_heat_flux} and \ref{fig:current_force} respectively. The Peltier heat flux  $\boldsymbol{q}_{e,(\parallel,\perp)}^u$ and Ettingshausen heat flux $\boldsymbol{q}_{e,\times}^u$ are nonlocal for Hall parameters $\chi_0 \lesssim 1$. Similarly to the nonlocal conductive heat flux, the peak heat flux is inhibited, and the heat flux is enhanced away from the peak current. The friction and cross friction are less sensitive to nonlocal effects than the heat fluxes since they are lower velocity moments, but still experience qualitatively similar deviations. Similarly to the temperature-gradient-driven fluxes, the cross components for moderate magnetizations are much more influenced by nonlocal behavior than their perpendicular counterparts. 

\subsection{Strongly magnetized nonlocal transport}

For large Hall parameters $\chi_e \gtrsim 10$, the plasma is sufficiently strongly magnetized that transport is local even for Knudsen numbers $N_K \sim 0.1$. Here, the Larmor radius $\rho_L= v_{\text{th},e} / \omega_{\text{c},e} $ becomes shorter than the gradient lengthscale $L$, and therefore becomes the limiting factor for nonlocal behavior. However, such a strongly magnetized regime then increases the maximum Knudsen number for which the fluid model is justified.
\begin{figure}[htpb!]
    \centering    
    \includegraphics[width=0.5\textwidth]{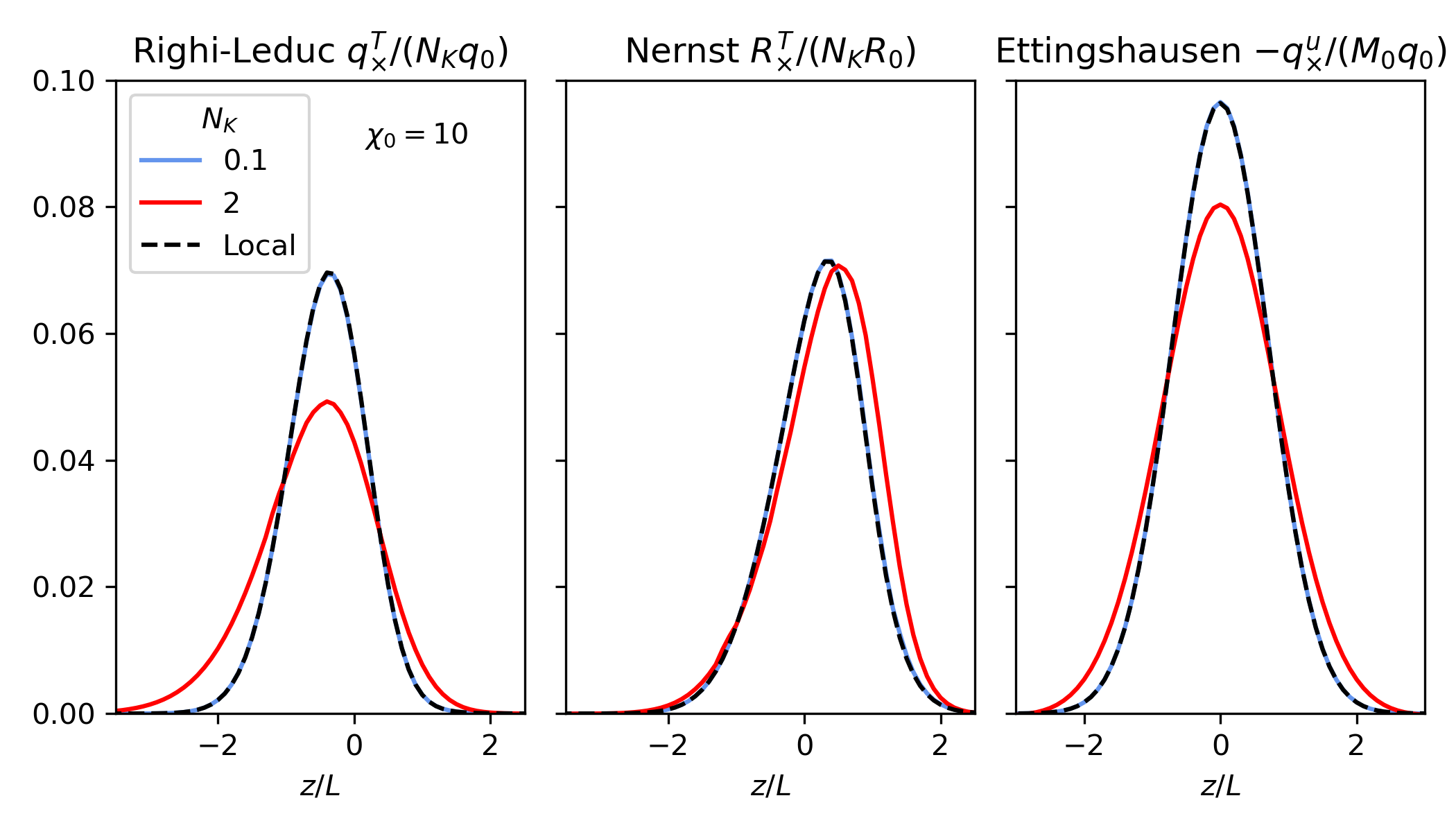}
    \caption{
        Righi-Leduc heat flow, Nernst force, and Ettingshausen heat flow for strongly magnetized plasma.
    }
    \label{fig:strongly_magnetized}
\end{figure}

Figure \ref{fig:strongly_magnetized} shows the temperature-gradient-driven and current-driven cases above in a strongly magnetized plasma. Only the cross terms are shown, since these are the dominant transport terms for high magnetization, assuming the thermodynamic drive ($\boldsymbol{\nabla}T_e$ or $\Delta \boldsymbol{u}$) is perpendicular to the magnetic field. For the already large Knudsen number $N_K=0.1$, the strong magnetic field completely suppresses nonlocal behavior, forcing the fluxes to their local results. The strong magnetic field then allows for a larger Knudsen number where a fluid description is valid, so here a Knudsen number $N_K=2$ is shown, where the gradient is sufficiently sharp that nonlocal effects emerge again. 

\section{Kinetics}\label{sec:kinetics}

\subsection{Nonlocal distribution function}

The \ac{RKM} naturally solves for the \ac{DF} $f_e$ from the spherical harmonic expansion \ref{spherical_harmonic_expansion}. Here, we plot the \ac{DF} $f_e$ for the nonlocal magnetized temperature gradient case above to visualize how nonlocality and magnetic fields distort the tail of the \ac{DF}.

\begin{figure}[htpb!]
    \centering    \includegraphics[width=0.5\textwidth]{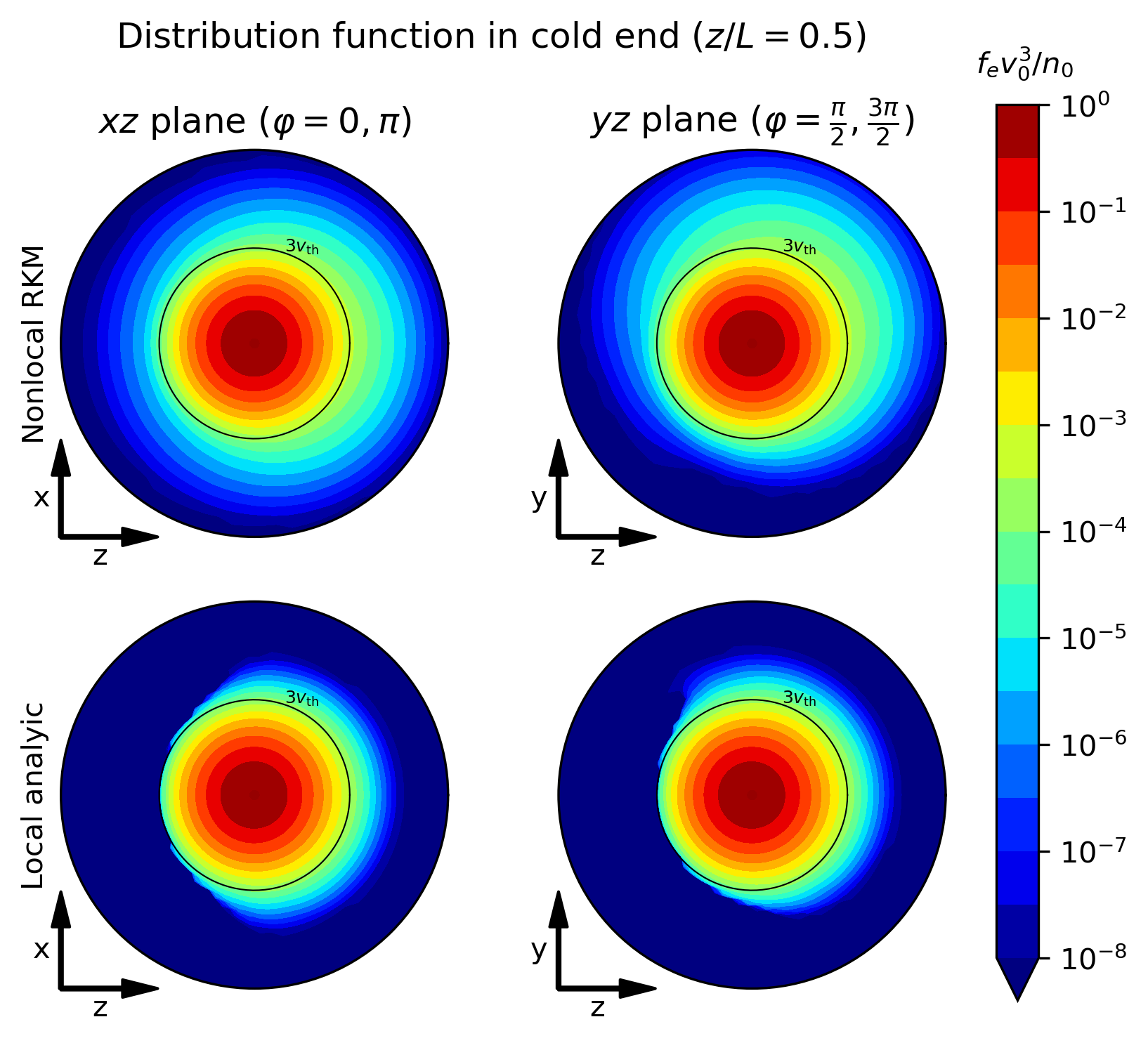}
    \caption{
        Electron \ac{DF} in the cold end for the $N_K=0.1$ magnetized ($\chi_0 = 0.5$) isochoric temperature step case above. A circle at $w=3v_{\text{th}}(z/L=0.5)$ is shown to indicate the radial velocity scale.
    }
    \label{fig:nonlocal_polar_df}
\end{figure}

Figure \ref{fig:nonlocal_polar_df} shows the \ac{DF} $f_e$ in the cold end ($z/L=0.5$) in the nonlocal magnetized case in Section \ref{subsection_temperature_gradient} above. In the $xz$ plane, containing the temperature gradient and magnetic field, the local solution is anisotropic in the $+\boldsymbol{\hat{z}}\parallel-\boldsymbol{\nabla}T_e$ direction. In the $yz$ plane, there is a deflection of this anisotropy in the $+\boldsymbol{\hat{y}}\parallel\boldsymbol{B}\times\boldsymbol{\nabla}T_e$ direction due to the magnetic field. Sufficiently far out in the tail at the rear of the distribution ($w/v_{\text{th}} \gtrsim 3 , \, \vartheta \gtrsim 3\pi/4$), the local solution is negative, highlighting the limit of its validity. The tail of the nonlocal distribution is enhanced beyond $w/v_{\text{th}} \gtrsim 3$ due to near-collisionless suprathermal particles streaming from the hot end. Due to the magnetic field, this suprathermal tail is deviated in the $+\boldsymbol{\hat{y}}\parallel\boldsymbol{B}\times\boldsymbol{\nabla}T_e$ direction. The nonlocal solution corrects the unphysical negative-ness of the \ac{DF} in the rear of the tail present in the local analytic solution. 

\subsection{Heat flux integrands}

To better understand nonlocal and magnetized effects on the conductive heat flow, we may examine the behavior of the heat flux integrand.

\begin{figure}[htpb!]
    \centering    \includegraphics[width=0.5\textwidth]{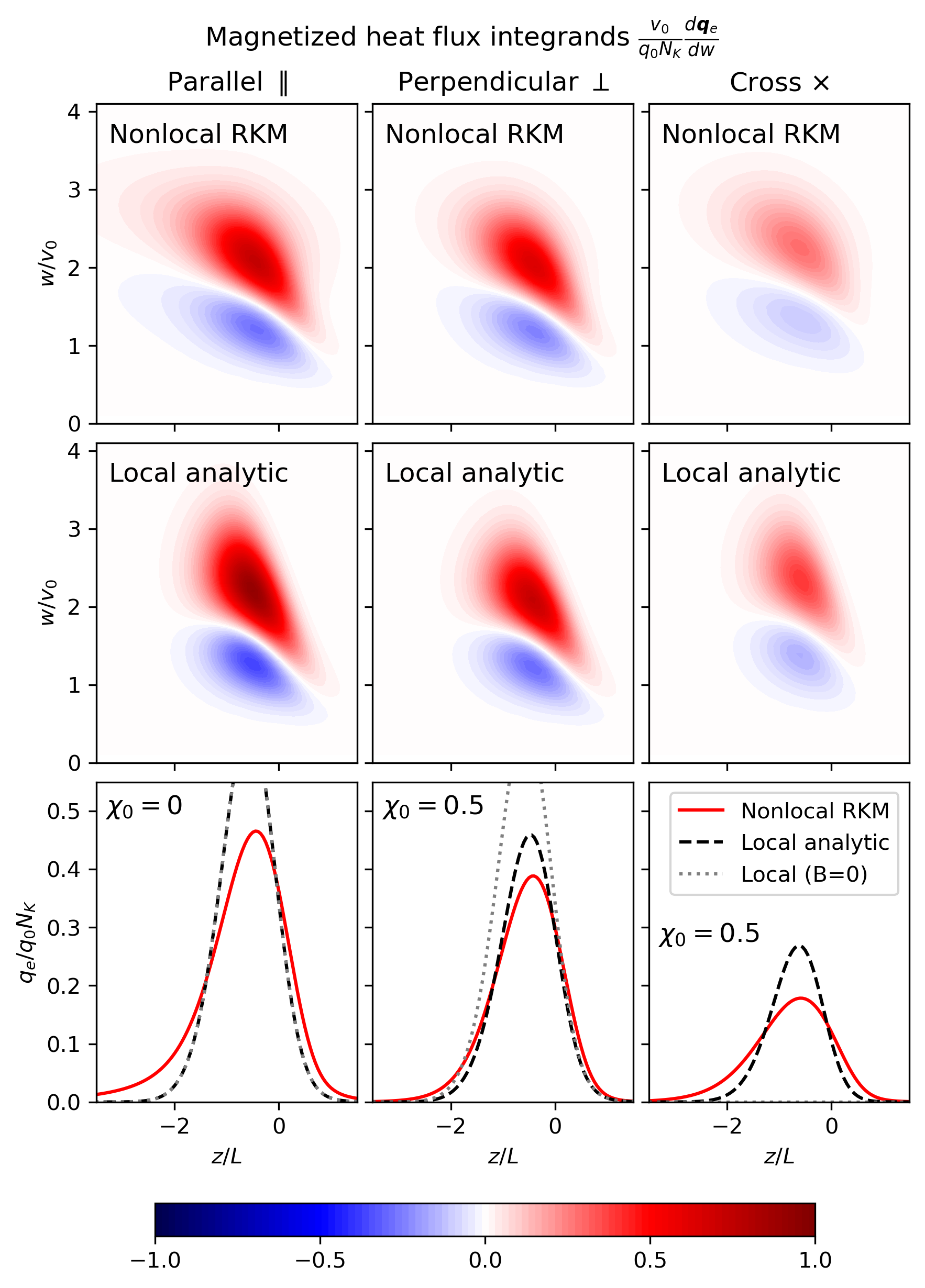}
    \caption{
        Integrand of conductive heat flux components for the unmagnetized and magnetized temperature gradient cases with $N_K=0.1$ obtained from the \ac{RKM} and local analytic result.
    }
    \label{fig:heat_flux_integrands}
\end{figure}

Figure \ref{fig:heat_flux_integrands} shows the components of the electron conductive heat flux integrand $\propto w^5 \boldsymbol{S}_{1,e}$ for the unmagnetized and magnetized temperature gradient cases presented above. In the unmagnetized case, nonlocal effects distort the heat flux integrand at higher velocities $w$ where \acp{MFP} are longer, smearing the integrand across the spatial domain. In contrast to the local analytic solution, a beam-like feature penetrating into the cold end for $w\gtrsim 2 v_0$ is present, giving rise to the preheat seen in the nonlocal heat flow below. Similar enhancement is also seen in the hot end, where nonlocal effects allow for suprathermal particles to stream across the gradient.

The magnetic field suppresses this nonlocal behavior at larger $w$, largely inhibiting the beam features and preheat seen in the cold end. Nonlocal distortions on the integrand and heat flux are still present, but less than the unmagnetized case. The magnetic field then adds a nontrivial component in the cross $\times$ direction, giving rise to the Righi-Leduc heat flow. The local Righi-Leduc integrand is more weighted towards higher $w$ than the parallel and perpendicular conductive components, therefore nonlocal effects then have a greater effect on the Righi-Leduc heat flow than the perpendicular component.

\subsection{Validity of collision operator}\label{a_posteriori_check}

The key assumption of the \ac{RKM} is neglecting the nonlinear component of the collision operator, which is the last term when expanding the collision operator
\begin{equation}\begin{aligned}
    C_{\alpha\beta}\{f_\alpha,f_\beta\} &= C_{\alpha\beta}\{f^M_\alpha,f^M_\beta\} + C_{\alpha\beta}\{ f^M_\alpha,\delta f_\beta\} \\&  + C_{\alpha\beta}\{ \delta f_\alpha,f^M_\beta\}+ C_{\alpha\beta}\{\delta f_\alpha,\delta f_\beta\},
\end{aligned}\end{equation}
where general species subscripts $\alpha$ and $\beta$ are reintroduced to emphasize that $C_{\alpha\beta}$ represents collisions of species $\alpha$ against species $\beta$. Focusing on the last two terms in the collision operator expansion, the test-particle and nonlinear components, we have
\begin{equation}\begin{aligned}
    C_{\alpha\beta}\{\delta f_\alpha , f^M_\beta\} + C_{\alpha\beta}\{ \delta f_\alpha , \delta f_\beta\} \equiv C_{\alpha\beta}\{ \delta f_\alpha , f_\beta \} 
\end{aligned}\end{equation}
Neglecting the nonlinear component compared to the test-particle operator is then equivalent to approximating the Rosenbluth potentials appearing in this operator as those evaluated from a Maxwellian, i.e. $\mathcal{H}_\beta \approx \mathcal{H}_\beta^M$ and $\mathcal{G}_\beta \approx \mathcal{G}_\beta^M$. This assumption is justified \textit{a posteriori} by checking the ratios of the deviation of the Rosenbluth potentials $\delta \mathcal{H}_\beta $ from their Maxwellian results $\mathcal{H}^M_\beta$ (and identically for $\mathcal{G}$) are small.
\begin{figure}[htpb!]
    \centering    \includegraphics[width=0.5\textwidth]{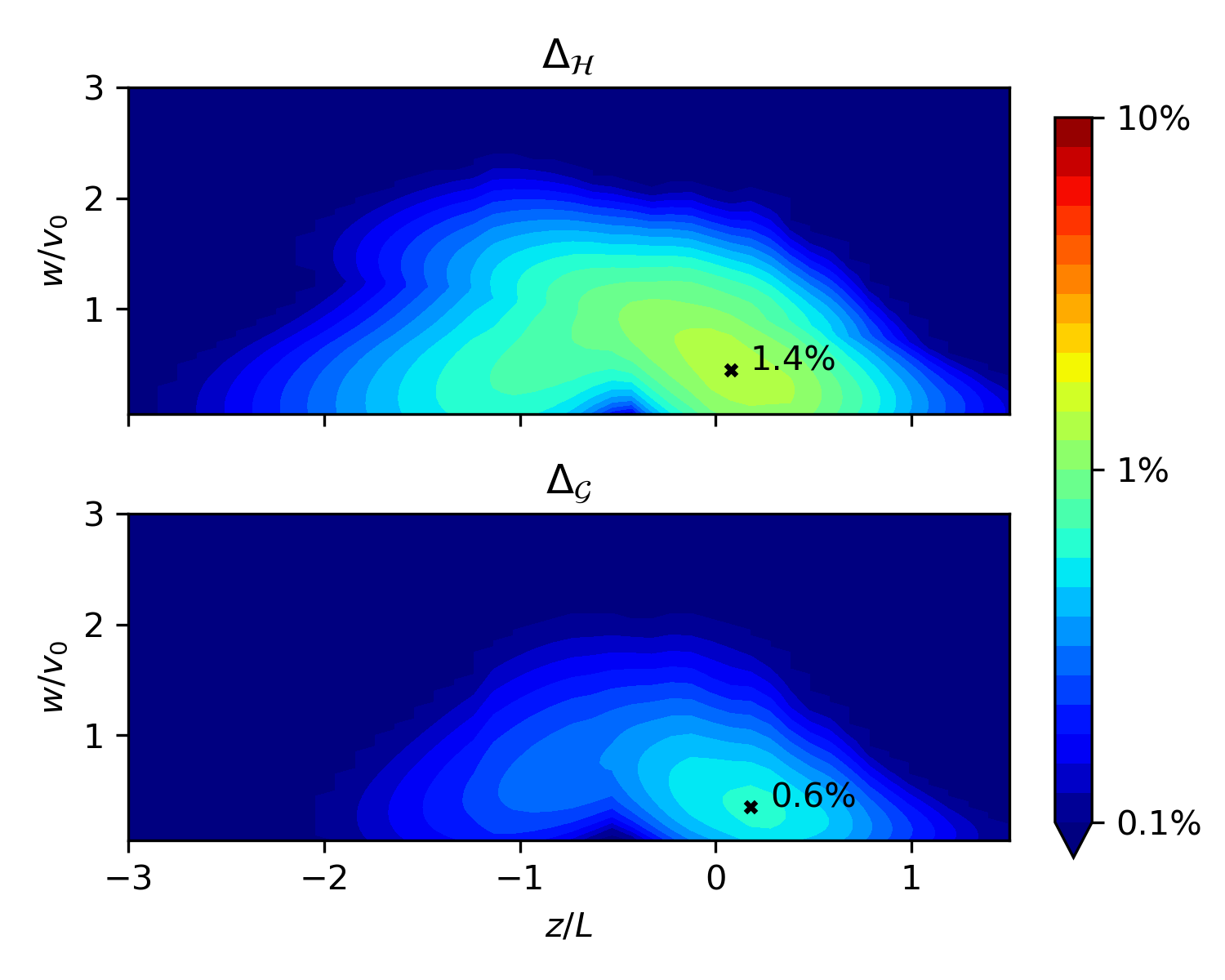}
    \caption{
        \textit{A posteriori} justification to the smallness of the nonlocal correction to the electron Rosenbluth potentials $\mathcal{H}$ and $\mathcal{G}$ for the case of an isochoric unmagnetized plasma with a temperature gradient of Knudsen number $N_K=1/10$. The smallness of these ratios ($\lesssim1\%$) across phase space justifies neglecting the nonlinear component of the collision operator for $N_K \lesssim 1/10$.
    }
    \label{fig:rosenbluth_potentials}
\end{figure}
Figure \ref{fig:rosenbluth_potentials} shows the deviation of the Rosenbluth potentials from their Maxwellian results for the unmagnetized temperature gradient case with $N_K=0.1$. The deviation of $\mathcal{H}$ is defined as
\begin{equation}\begin{aligned}
    \Delta_{\mathcal{H}} \equiv \frac{1}{\vert\mathcal{H}^{M}\vert} \sum_{l=0}^{N_l-1}\sum_{m=-l}^l  \frac{2l+1}{4\pi}\frac{(l-m)!}{(l+m)!} \vert \delta \mathcal{H}^{(l,m)} \vert,
\end{aligned}\end{equation}
which is an upper bound for $\vert \delta \mathcal{H} / \mathcal{H^M} \vert \leq \Delta_{\mathcal{H}}$ over velocity angle space, and similarly for $\mathcal{G}$. The maximum deviations are $\Delta_{\mathcal{H}}^{\text{max}}\approx 1.4\%$ and $\Delta_{\mathcal{G}}^{\text{max}}\approx 0.6\%$, justifying the assumption of neglecting the nonlinear component of the collision operator for Knudsen numbers $N_K \lesssim 1/10$. 

\subsection{Solving for nonlocal deviation}

The \ac{DDF} equation may be written as
\begin{equation}\label{electron_DDF_equation_T_and_F}
    D = (\hat{C}^T + \hat{C}^F - \hat{V} - \hat{B}) \delta f,
\end{equation}
where test-particle and field-particle operators are written separately and species subscripts are dropped for brevity. The local leading-order Chapman-Enskog solution for the \ac{DDF} is obtained by solving Equation \ref{electron_DDF_equation_T_and_F} to leading order in Knudsen number, which then becomes
\begin{equation}\label{local_electron_DDF_equation}
    D^{(0)} = (\hat{C}^T + \hat{C}^F - \hat{B}) \delta f^{\text{local}}.
\end{equation}
If the \ac{DDF} is decomposed via $\delta f =\delta f^{\text{local}} + g $, then the equation for the nonlocal correction $g$ is obtained by subtracting Equation \ref{local_electron_DDF_equation} from Equation \ref{electron_DDF_equation_T_and_F} to yield
\begin{equation}\label{electron_g_equation}
\begin{aligned}
    D'= (\hat{C}^T + \hat{C}^F -\hat{V} - \hat{B})  g,
\end{aligned}  
\end{equation}
where the new inhomogeneous driving term $D'\equiv D - D^{(0)} + \hat{V}\delta f^{\text{local}}\approx \hat{V}\delta f^{\text{local}} $. Equation \ref{electron_g_equation} appears to be just as complex as the original \ac{DDF} equation, only with a different inhomogeneous term on the left hand side. However, there is additional computational benefit to this decomposition. The test-particle collision operator is a differential operator in velocity space, whereas the field-particle operator is an integral operator in velocity space. When discretizing, the test-particle operator is then sparse in the velocity index, only linking nearest neighbors $j\pm1$, whereas the field-particle operator is dense in the velocity index, linking all $j$ for given spatial position $i$ and harmonic $l,m$. The \ac{RKM} is still tractable with the field-particle operator included, but would be more computationally efficient without it. Discarding the field-particle operator from the \ac{DDF} equation is unjustified in a first-principles description, and doing so means that the local limit is no longer reproduced. However, when solving for the nonlocal correction $g$, dropping the field-particle operator may be justified since the field-particle operator is only significant in the bulk of the distribution, whereas nonlocal effects are most significant in the tail. Discarding $\hat{C}^Fg$ leaves
\begin{equation}\begin{aligned}\label{g_equation}
    D' \approx (\hat{C}^T  -\hat{V} - \hat{B})  g,
\end{aligned}\end{equation}
which reproduces the local limit by construction since $g$ will be subleading to $\delta f^{\text{local}}$ for velocities relevant to transport for sufficiently small $N_K \lesssim 1/100$. 
\begin{figure}[htpb!]
    \centering    \includegraphics[width=0.5\textwidth]{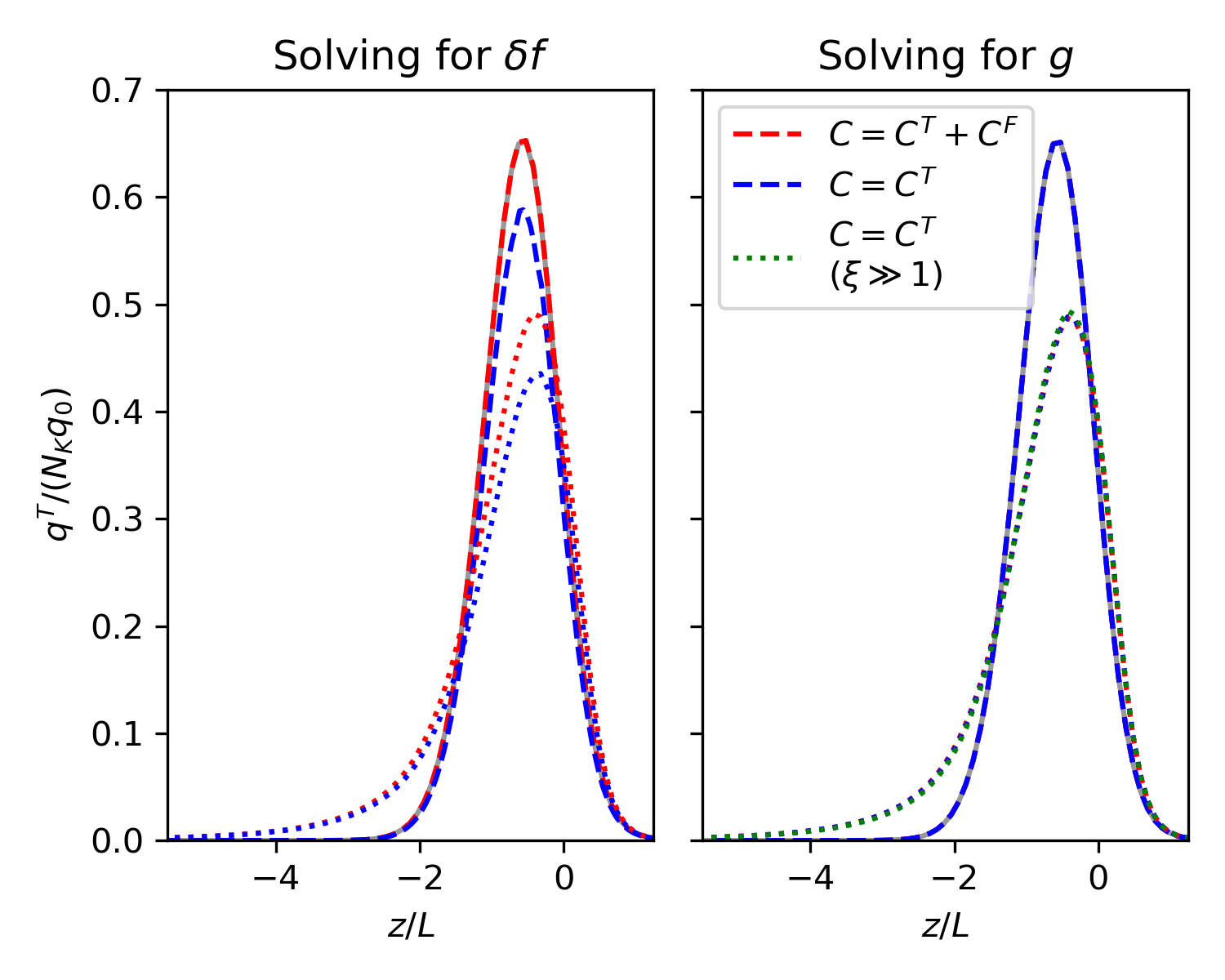}
    \caption{
        Comparison of unmagnetized nonlocal conductive heat flux for \ac{RKM} solving for $g = f - f^M - \delta f^{\text{local}}$ versus $\delta f = f - f^M$ with and without the field-particle operator included for $N_K = 0.001$ (dashed lines) and $N_K = 0.1$ (dotted lines).
    }
    \label{fig:solving_for_g}
\end{figure}

Figure \ref{fig:solving_for_g} compares the approaches of solving for $\delta f$ and $g$ with and without the field-particle operator included. When solving for $\delta f$, discarding the field-particle operator yields an incorrect result for the local and nonlocal heat flow. The local result has a thermal conductivity reduced by $\sim 10\%$, and the peak nonlocal heat flow is also reduced. However, the preheat of the nonlocal heat flow still matches well with the full result, suggesting that nonlocal heat flow is governed primarily by test-particle collisions. When solving for $g$, the local limit is reproduced by construction. The nonlocal heat flow obtained from $g$ is practically unchanged when the field-particle operator is discarded, leaving this as an even more tractable first-principles approach.

Furthermore, since the nonlocal deviation $g$ is largest in the tail of the distribution where $\xi \equiv w/v_{\text{th},e} \gtrsim 1$, we may use the asymptotic limits for the velocity-dependent collision frequencies appearing in the test-particle operator. This leaves the operator in the simpler form
\begin{equation}\begin{aligned}
    \hat{C}^T g \approx \hat{\nu}_{ee}(\boldsymbol{x}) \bigg[ - \frac{1+Z}{2\xi^3} \hat{\mathcal{L}}_\mu + \frac{1}{\xi^2}\partial_\xi\bigg( 1 + \frac{1}{2\xi} \partial_\xi  \bigg) \bigg) \bigg]g.
\end{aligned}\end{equation}
The resulting nonlocal heat flux from this approximation is shown with the leading order $\xi\gg1$ asymptotic expansions for $\nu_D$, $\nu_s$, and $\nu_\parallel$, which is still in very good agreement with the original result. Equation \ref{g_equation} with these asymptotic forms for the collision frequencies is then an especially simple form of the first-principles \ac{RKM} which could be tractable in multiple spatial dimensions. It has been verified that dropping any of the test-particle terms originating from $\nu_D$, $\nu_s$, or $\nu_\parallel$ strongly disturbs the nonlocal solution, even if only done for $l=0$ or $l\geq 1$ terms. The test-particle operator may therefore not be simplified further by dropping such terms.

\section{Ion transport}\label{sec:ion_results}

Similarly to electrons, ions also have transport quantities such as heat flux and the viscous stress tensor. Nonlocal transport studies focus almost entirely on the the electron heat flux since electron heat flux is expected to dominate the ion heat flux since electrons are much lighter than ions. However, magnetized electron transport $\chi_e \gtrsim 10$, hotter ions $T_i> T_e$, and sharper ion temperature gradients $ \vert \boldsymbol{\nabla}T_i \vert > \vert \boldsymbol{\nabla}T_e \vert$ all allow for ion heat flow to become comparable to or dominate electron heat flow. Single-species ion heat flow is qualitatively similar to electron heat flow, therefore is not shown here since electron heat flow has already been clearly presented in the electron case. Viscous effects, in contrast, are almost always dominated by ions, therefore are presented here for ions rather than electrons. It should be noted that ion-electron collisions significantly modify ion transport coefficients \cite{10.1063/1.4922755} and therefore are also expected to impact nonlocal ion heat flow. These effects are readily included in the \ac{RKM} and will be presented in future work.


\subsection{RKM for ions}

The \ac{RKM} may be reformulated for a single ion species in a simple plasma with \ac{VFP} equation
\begin{equation}
\begin{aligned}\label{ion_VFP}
    \partial_t f_i &+ \boldsymbol{v}\cdot \boldsymbol{\nabla} f_i + \frac{Ze}{m_i}\big(\boldsymbol{E} + \boldsymbol{v}\times\boldsymbol{B}\big)\cdot\boldsymbol{\nabla}_{\boldsymbol{v}}f_i\\& = C_{ii}\{f_i,f_i\} + C_{ie}\{f_i,f_e\},
\end{aligned}
\end{equation}
with a similar derivation to before leading to the equation for the ion \ac{DDF}
\begin{equation}\label{ion_DDF_equation}
    D_i = \big( \hat{C}_{i} - \hat{V}_i - \hat{B}_i \big)\delta f_i,
\end{equation}
with the inhomogeneous driving term
\begin{equation}
\begin{aligned}
     &D_i = \bigg[ \bigg( \frac{w^2}{v_{\text{th},i}^2} - \frac{3}{2} \bigg)  K_i   + \frac{2}{v_{\text{th},i}^2}  \boldsymbol{\nabla} \boldsymbol{u}_i \colon \bigg( \boldsymbol{w}\boldsymbol{w} - \frac{1}{3}w^2\mathbb{I} \bigg) \\& +   \bigg( \bigg( \frac{w^2}{v_{\text{th},i}^2} - \frac{5}{2}  \bigg)  \boldsymbol{\nabla} \ln T_i - \frac{1}{p_i}\boldsymbol{\nabla}\cdot \Pi_i  \bigg) \cdot \boldsymbol{w}  \bigg] f^M_i ,
    \\&\text{where}\quad  K_i = \frac{2}{3 p_i}\bigg[ -\boldsymbol{\nabla}\cdot\boldsymbol{q}_i   - \Pi_i\colon \boldsymbol{\nabla}\boldsymbol{u}_i  \bigg],
\end{aligned}
\end{equation}
and operators
\begin{equation}
\begin{aligned}
    \hat{V}_i &= (\boldsymbol{w}+\boldsymbol{u}_i)\cdot \boldsymbol{\nabla} \\&- \bigg[ \boldsymbol{w}\cdot  (\boldsymbol{\nabla} \boldsymbol{u}_i)  + \frac{1}{\rho_i}\bigg( -\boldsymbol{\nabla}p_i -\boldsymbol{\nabla}\cdot \Pi_i  \bigg) \bigg] \cdot\boldsymbol{\nabla}_{\boldsymbol{w}},
    \\  \hat{B}_i &= \frac{Z e}{m_i}(\boldsymbol{w}\times\boldsymbol{B})\cdot\boldsymbol{\nabla}_{\boldsymbol{w}}.
\end{aligned}
\end{equation}
The ion-electron Maxwellian collision terms $C_{ei}\{f^M_i,f^M_e\}$ have canceled exactly with the $\boldsymbol{R}_e$ and $\tilde{W}_{ie}$ terms in $D_i$, and similarly the $\boldsymbol{R}_e$ term in $C_{ie}\{\delta f_i,f^M_e\}$ and $\hat{V}_i$ have cancelled. These cancellations mean that all electron transport terms have vanished from the ion \ac{DDF} equation so the ion solution does not require solving for electron transport first. The only appearance of electrons in the ion \ac{DDF} equation is through the ion-electron test-particle collision operator $\hat{C}_{ie} \delta f_i $. The ion \ac{DDF} equation is then projected onto the spherical harmonic basis, yielding
\begin{equation}
    \mathcal{D}_i^{(l,m)}  = \mathcal{C}_{i}^{(l,m)}  - \mathcal{V}_{i}^{(l,m)} - \mathcal{B}_{i}^{(l,m)},
\end{equation}
which is solved similarly to the electron case.

\subsection{Viscous stress tensor}

The viscous stress tensor represents the flux of momentum due to gradients in the flow. Unlike heat flux, where electron heat flux dominates ion heat flux unless electrons are magnetized or there is a significant temperature separation, ion viscosity is generally larger than or dominates over its electron counterpart. To the authors' knowledge, no full kinetic studies or closures studying nonlocality of the viscous stress tensor exist in previous literature, even without a first-principles approach or without magnetic fields. With the \ac{RKM}, the viscous stress tensor can be naturally recovered by integrating over the \ac{DDF} as specified in Equation \ref{transport_from_lm}.

While an in-depth investigation of nonlocal viscosity is beyond the scope of this publication, our preliminary study with the \ac{RKM} has demonstrated many novel nonlocal features of the viscous stress tensor. Firstly, there are two dimensionless numbers that impact its nonlocal behavior: the Mach number $M \sim \vert \boldsymbol{u}_i \vert / v_{\text{th},i}$ which governs the characteristic speed of the flow, and the Knudsen number $N_K \sim \lambda_{\text{th},ii} \vert \boldsymbol{\nabla} \ln( \vert\boldsymbol{u}_i\vert) \vert^{-1} $ which governs the length scale over which the flow profile varies. When $N_K \lll 1$ and $M \lesssim \mathcal{O}(1)$, the viscous stress tensor is local. However, when $N_K \gtrsim 1/10$, the viscous stress tensor becomes nonlocal, and in this regime the Mach number $M \sim \mathcal{O}(1)$ has further nonlocal effects on the stress tensor through the $\hat{V}_i \delta f_i$ term. This work takes a fixed $M \sim \mathcal{O}(1)$. 

All of the transport quantities studied above are rank-one tensors, or vectors, which are evaluated from the $l=1$ components of the spherical harmonic expansion. In contrast, viscous stress is represented by a rank-two tensor evaluated from the $l=2$ components. Other nonlocal closures either do not include the $l\geq 2$ terms or use a minimum entropy closure for the $l=2$ component, therefore are unlikely to correctly capture the nonlocal behavior of the viscous stress tensor which is directly evaluated from the $l=2$ component. The \ac{RKM} expands the \ac{DF} to an arbitrary order of spherical harmonics, with $N_l=5$ found to provide good convergence for this work.

There are two types of flows that generate stress: compressive flows where the flow varies in the same direction as the flow $\boldsymbol{u}_i \parallel \boldsymbol{\nabla}$ such as $\boldsymbol{u}_i=u_z(z)\boldsymbol{\hat{z}}$, and shear flows where the flow varies in a direction perpendicular to the flow $\boldsymbol{u}_i \perp \boldsymbol{\nabla}$ such as $\boldsymbol{u}_i = u_y(z)\boldsymbol{\hat{y}}$. For this work, we present results for a shear flow with a uniform magnetic field $\boldsymbol{B}=-B_0 \boldsymbol{\hat{x}}$.
\begin{figure}[htpb!]
    \centering   \includegraphics[width=0.5\textwidth]{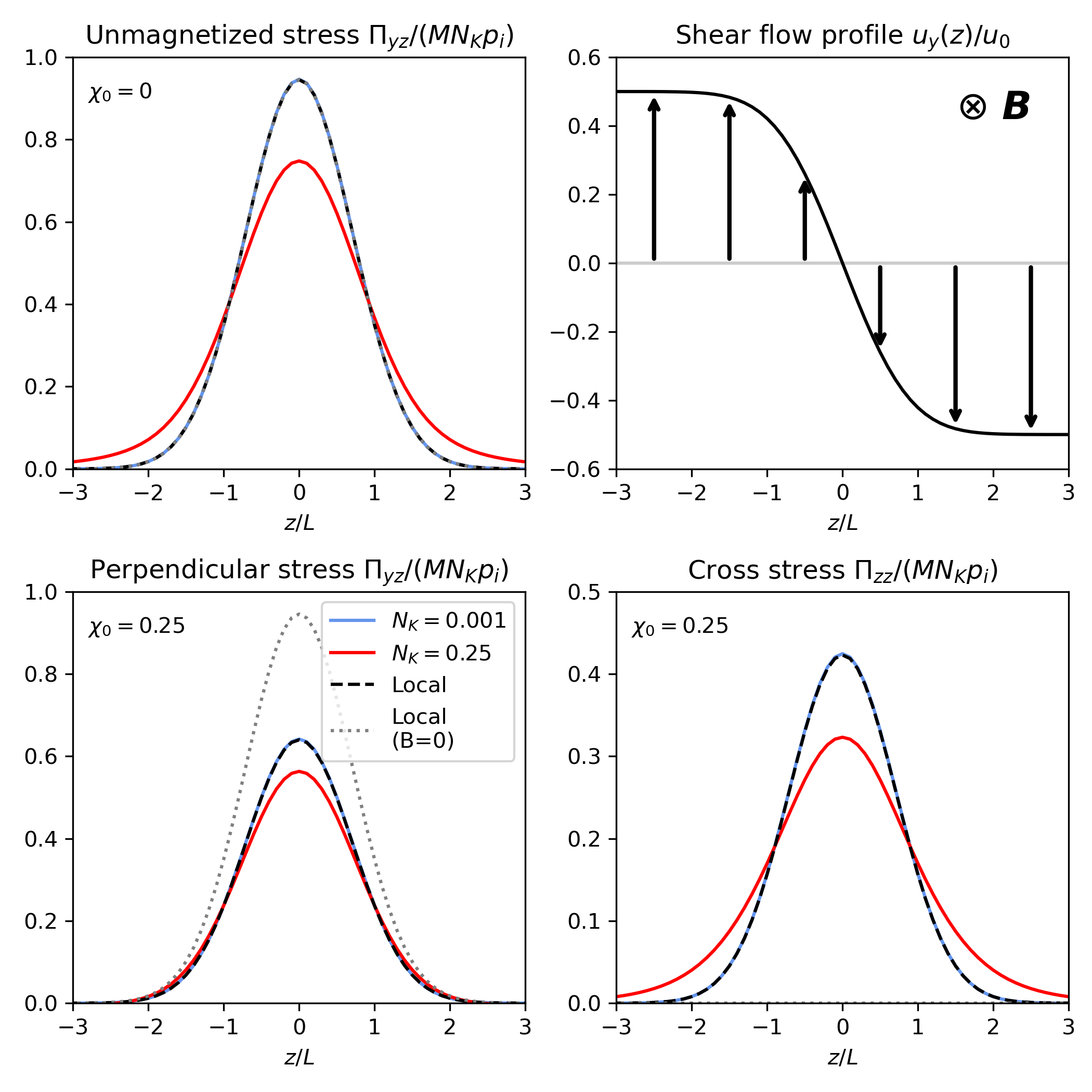}
    \caption{
       Nonlocal ion viscous stress tensor for a shear ion flow $\boldsymbol{u}_i = - \frac{1}{2}u_0\,\text{erf}(z/L)\boldsymbol{\hat{y}}$ with Mach number $M = \vert u_0 \vert/ v_{\text{th},i} = 1$. The stresses are normalized against the Mach number $M$, Knudsen number $N_K$, and ion pressure $p_i=n_i T_i $.
    }
    \label{fig:nonlocal_viscous_stress}
\end{figure}
Figure \ref{fig:nonlocal_viscous_stress} shows the viscous stress components $\Pi_{yz}$ and $\Pi_{zz}$ generated by a shear flow $\boldsymbol{u}_i = -\frac{1}{2}u_0\,\text{erf}(z/L)\boldsymbol{\hat{y}}$ in an isothermal isochoric simple hydrogen plasma. Similarly to results for vectorial transport quantities, the \ac{RKM} accurately reproduces local analytic results for the viscous stress components for small Knudsen number. For large Knudsen number $N_K=0.25$, the unmagnetized parallel stress $\Pi_{yz}$ is significantly modified by nonlocal effects with qualitatively similar behavior as other transport quantities; the peak flux is inhibited, and a `preheat' is developed, where the flux away from the main gradient is enhanced due to streaming suprathermal particles. In the magnetized case $\chi_{i,0} = 0.25$, the magnetic field restores the nonlocal perpendicular stress closer to its local result. The cross stress $\Pi_{zz}$ appears more strongly influenced by nonlocal effects than the magnetized shear stress, analogously to the Righi-Leduc heat flux $\boldsymbol{q}^T_{\alpha,\times}$ being more sensitive to nonlocality than the perpendicular counterpart $\boldsymbol{q}^T_{\alpha,\perp}$. 

\section{Discussion}\label{section_conclusion}

Magnetoinertial schemes such as MagLIF \cite{slutz_high-gain_2012,Yager-Elorriaga_2022} are a promising approach to confined fusion for energy production . These approaches apply external magnetic fields to suppress electron heat conduction or use the $\boldsymbol{j}\times\boldsymbol{B}$ to compress cylindrical liners to approach the high densities and temperatures required to achieve thermonuclear burn. Even in unmagnetized \ac{ICF} implosions, the Biermann battery effect can spontaneously generate magnetic fields strong enough to magnetize the plasma \cite{PhysRevLett.118.155001}. Correct modeling of magnetized transport within simulations of \ac{ICF} implosions is therefore increasingly important for the path towards inertial fusion energy. 

A first-principles \ac{RKM} has been extended from previous work to include magnetic fields. The \ac{RKM} solves a linear form of the \ac{VFP} equation with the Fokker-Planck collision operator. The \ac{RKM} is shown to reproduce all classical (Braginskii-like) transport terms in the short mean free path cases. In the nonlocal cases, the implemented collision operator has been validated \textit{a posteriori}, with Rosenbluth potentials deviating by $\lesssim 1\%$ from their Maxwellian forms for $N_K \lesssim 0.1 $.


An alternative formulation of the \ac{RKM} is also presented, solving for the nonlocal deviation of the \ac{DF} $g$ rather than the deviation from Maxwellian $\delta f$. The idea behind this approach can be summarized as follows: if the field-particle operator is discarded when solving for $\delta f$, both local and nonlocal transport results are noticeably different from their nearly exact counterparts obtained with the full $\delta f$ method of the previous paragraph. However, when solving for the nonlocal deviation, both local and nonlocal transport results are found to be unchanged if the field-particle operator is discarded, therefore neglecting it is well justified. The resulting equation for $g$, with the test-particle collision operator, is more computationally tractable and well suited to act as a nonlocal transport closure. 

The \ac{RKM} may be applied as a nonlocal transport closure within a fluid code to investigate the impact of nonlocal transport on hydrodynamic evolution. The \ac{RKM} could also be used to train a surrogate nonlocal closure based on machine learning techniques for an especially computationally efficient approach, or be used to study other kinetic phenomena besides transport such as reactivity and neutron spectra subject to Knudsen-layer effects. 

The standard nonlocal behavior of the unmagnetized electron conductive heat flux is demonstrated, where a sharp temperature gradient results in a nonlocal heat flux with an inhibited peak flux and preheats away from the main gradient. Introducing a perpendicular magnetic field suppresses nonlocal effects on the conductive heat flux along the temperature gradient. The magnetic field also introduce the Righi-Leduc heat flow, which is more sensitive to nonlocal effects than its perpendicular counterpart. Similarly, nonlocal effects are also present on the thermal force and Nernst term.

As well as considering a temperature gradient to investigate nonlocality of temperature-gradient-driven transport, a current profile of different length scales is also studied. These cases demonstrate similar nonlocal effects on current-driven transport, namely the Peltier and Ettingshausen heat flows and electron-ion friction and cross friction. The \ac{RKM} has also been applied to demonstrate more non-standard nonlocal behavior such as the enhancement of current-driven heat flow by large currents \cite{mitchell_nonlocal_2025}. 

Strong magnetic fields $\chi \gg 1$ suppress nonlocal behavior, restoring transport fluxes back to their magnetized local results. For strong magnetization, it is typically the `cross' components such as the Righi-Leduc heat flow that dominate over their perpendicular counterparts. At sufficiently high Knudsen numbers, nonlocal effects then reemerge on these cross components.

The \ac{RKM} expands up to an arbitrary order in spherical harmonics, therefore the viscous stress tensor is naturally included in this description. The ion rather than electron viscous stress tensor is presented, since ion viscous effects almost always dominate over electron viscous effects. Nonlocal behavior of the ion viscous stress tensor appears due to sharp shearing ion flows. A magnetic field then suppresses nonlocal effects of the ordinary viscous stress tensor, but also introduces a gyroviscous component which, analogously to the Righi-Leduc heat flow, is more sensitive to these nonlocal effects. In existing literature, there appears to be no other studies of nonlocal viscosity, even for unmagnetized plasmas.

Full kinetic simulations usually serve as numerical experiments for investigating nonlocal transport rather than closures. Consider using a full \ac{VFP} simulation to study nonlocal transport in the temperature gradient case of Section \ref{subsection_temperature_gradient}. The \ac{DF} would be initialized as a Maxwellian distribution with a temperature varying across space. When the simulation is run, the \ac{DF} evolves to gain an anisotropic component, which relaxes over a few collision times to a quasi-stationary result. For initial profiles with low Knudsen numbers, this quasi-stationary result should match the local Chapman-Enskog solution at each point in space. For large initial Knudsen numbers, the quasi-stationary solution will include nonlocal effects of interest; however, the temperature gradient will partially relax on a similar timescale to the initial relaxation to the quasi-stationary result. The heat flow in this nonlocal case then corresponds to the partially relaxed temperature profile rather than the initial temperature profile. It is then difficult for a full \ac{VFP} code to provide a robust comparison between different Knudsen numbers for the same shape of profile, a more detailed parameter scan over different temperature profiles, or as a transport closure to a fluid code. In addition, the \ac{DF} will also gain some nonzero flow profile during the initial relaxation period. This flow should itself drive heat flow, which cannot be distinguished from the conductive heat flow when integrating over the \ac{DF}. Some approaches may counteract this by adding sources or sinks of particle number, momentum, or energy to constrain the \ac{DF} to have the same density, flow, and temperature throughout the initial relaxation period. However, it is not clear if these constraints themselves modify the physics of the heat flow. The initial Maxwellian condition can also lead to strange behavior of the simulation at early times such as oscillating heat flow \cite{10.1063/5.0086783,dearling_transport_2024} and therefore cooling of cold regions. Since the \ac{RKM} evaluates the \ac{DF} from the instantaneous fluid profiles, it does not have this issue when finding nonlocal heat flow from hydrodynamic profiles. The \ac{RKM} could instead be used to initialize the \ac{DF} accounting for nonlocal effects in full \ac{VFP} codes to avoid this strange behavior.

Full kinetic approaches are integrated simulations which do not allow for different nonlocal effects to be clearly studied in isolation, in contrast to the \ac{RKM} where these effects may be isolated by removing the appropriate terms from the \ac{DF} equation when solving. For example, a full \ac{VFP} code is not able to separate out heat flow contributions from temperature gradients versus currents. In the \ac{RKM}, these correspond to two distinct thermodynamic driving forces, therefore the conductive and current-driven heat flows can be found independently for the same hydrodynamic profiles. This provides a clearer decoupled insight into how different parts of the input hydrodynamic profiles give rise to different nonlocal features.






\begin{strip}

\appendix

\section{Spherical Harmonic Projections}\label{appendix_projection}

Similarly to previous work, the projections of the linear collision operators are
\begin{equation}
\begin{aligned}
     & \mathcal{C}_{\alpha\beta}^{T,(l,m)} = - \frac{1}{2}l(l+1) \nu_D^{\alpha\beta} \delta f_\alpha^{(l,m)} + \frac{1}{w^2} \partial_w \bigg( \tilde{\nu}_s^{\alpha\beta} w^3\delta f_\alpha^{(l,m)} + \frac{1}{2}\nu_\parallel^{\alpha\beta} w^4 \partial_w\delta f_\alpha^{(l,m)} \bigg),
    \\ & \mathcal{C}_{\alpha\beta}^{F,(l,m)} =\bigg(\frac{q_\alpha q_\beta}{m_\alpha \varepsilon_0} \bigg)^2 \ln \Lambda_{\alpha\beta} \frac{1}{v_{\text{th},\alpha}^{2}}  f^M_\alpha  \Biggl[ \frac{m_\alpha}{m_\beta} v_{\text{th},\alpha}^2 \delta f_\beta^{(l,m)} - \frac{l(l-1)}{2l-1} \frac{1}{v_{\text{th},\alpha}^2} \Psi_{\beta,3}^{(l,m)}(w) +  \frac{(l+1)(l+2)}{2l+3} \frac{1}{v_{\text{th},\alpha}^2} \Psi_{\beta,4}^{(l,m)}(w)   \\&  + \bigg( - 1 + \bigg(\frac{m_\alpha}{m_\beta} - 1 \bigg) l + \frac{(l+1)(l+2)}{2l+3} \frac{ w^2}{v_{\text{th},\alpha}^2} \bigg) \Psi_{\beta,1}^{(l,m)}(w)
     + \bigg( - 1 - \bigg(\frac{m_\alpha}{m_\beta} - 1 \bigg) (l+1) - \frac{l(l-1)}{2l-1} \frac{ w^2}{v_{\text{th},\alpha}^2}   \bigg) \Psi_{\beta,2}^{(l,m)}(w)
    \Biggl],
\end{aligned}
\end{equation}
where we have used that spherical harmonics are eigenfunctions of the pitch-angle scattering operator via $\hat{\mathcal{L}}(P_l^me^{im\varphi}) = -l(l+1)P_l^me^{im\varphi}$ and defined the integrals
\begin{equation}\begin{aligned}\label{Psi_integrals}
    & \Psi_{\beta,1}^{(l,m)}(w) = \frac{2}{2l+1} w^l \int_w^\infty du\ u^{1-l} \delta f_\beta^{(l,m)}(u), \quad \Psi_{\beta,2}^{(l,m)}(w) = \frac{2}{2l+1} \frac{1}{w^{l+1}} \int_0^w du\ u^{2+l} \delta f_\beta^{(l,m)}(u),
    \\& \Psi_{\beta,3}^{(l,m)}(w) = \frac{2}{2l+1} w^l \int_w^\infty du\ u^{3-l} \delta f_\beta^{(l,m)}(u),\quad  \Psi_{\beta,4}^{(l,m)}(w) = \frac{2}{2l+1} \frac{1}{w^{l+1}} \int_0^w du\ u^{4+l} \delta f_\beta^{(l,m)}(u).
\end{aligned}\end{equation} 

The projection of the driving term is

\begin{equation}
\begin{aligned}
    \mathcal{D}_\alpha^{(l,m)} & = \int d^2\Omega\ e^{-im\varphi}P_l^m(\mu) D_\alpha
    = \bigg[ d_\alpha^0 \delta_{l,0}\delta^{m,0} + \boldsymbol{d}_{\alpha}^1 \cdot \langle l,m\vert \boldsymbol{\hat{w}} \rangle + \boldsymbol{d}_{\alpha,z}^2 \cdot \langle l,m\vert  \Big(\boldsymbol{\hat{w}} \boldsymbol{\hat{w}} - \frac{1}{3} \mathbb{I}\Big)\cdot\boldsymbol{\hat{z}}  \rangle \bigg] f^M_\alpha,
\end{aligned}
\end{equation}
with components of the driving force
\begin{equation}
\begin{aligned}
    d_{\alpha}^0 = \bigg( \frac{w^2}{v_{\text{th},\alpha}^2} - \frac{3}{2} \bigg)  K_\alpha
    ,\quad \boldsymbol{d}_e^1 &= w \bigg( \bigg( \frac{w^2}{v_{\text{th},e}^2} - \frac{5}{2}  \bigg)  \boldsymbol{\nabla}  \ln T_e + \frac{1}{p_e}\bigg( -\boldsymbol{\nabla} \cdot \Pi_e  + \boldsymbol{R}_e\bigg) + 2\hat{\nu}_{ei}  \frac{v_{\text{th},e}}{w^3}\Delta\boldsymbol{u}\bigg),
    \\ \boldsymbol{d}_{\alpha,z}^2 = 2\frac{ w^2 }{v_{\text{th},\alpha}^2}\boldsymbol{u}_\alpha', \quad \boldsymbol{d}_i^1 &= w \bigg( \bigg( \frac{w^2}{v_{\text{th},i}^2} - \frac{5}{2}  \bigg)  \boldsymbol{\nabla} \ln T_i -\frac{1}{p_i}\boldsymbol{\nabla}\cdot \Pi_i   \bigg),
\end{aligned}
\end{equation}
where $\boldsymbol{u}_\alpha' = \partial_z \boldsymbol{u}_\alpha $ and

\begin{equation}
\begin{aligned}
    \langle l,m\vert \boldsymbol{\hat{w}}\rangle &= \frac{2\pi}{3} \bigg[ - 2 \delta^{m,1}  (\boldsymbol{\hat{x}}-i\boldsymbol{\hat{y}})  +   \delta^{m,-1}(\boldsymbol{\hat{x}}+i\boldsymbol{\hat{y}})  + 2 \delta^{m,0} \boldsymbol{\hat{z}} \bigg] \delta_{l,1},
    \\ \langle l,m\vert \Big(\boldsymbol{\hat{w}} \boldsymbol{\hat{w}} - \frac{1}{3} \mathbb{I}\Big)\cdot\boldsymbol{\hat{z}} \rangle &= \frac{4\pi}{15} \bigg[ - 6 \delta^{m,1}  (\boldsymbol{\hat{x}}-i\boldsymbol{\hat{y}})  +   \delta^{m,-1}(\boldsymbol{\hat{x}}+i\boldsymbol{\hat{y}})  + 4 \delta^{m,0} \boldsymbol{\hat{z}} \bigg] \delta_{l,2}.
\end{aligned}
\end{equation}
The projection of the Vlasov and magnetic field terms are

\begin{equation}
\begin{aligned}
    \mathcal{V}_\alpha^{(l,m)} 
    & = u_{\alpha,z} \partial_z \delta f_\alpha^{(l,m)} + w \partial_z \langle l,m \vert \mu \delta f_\alpha \rangle -  w\boldsymbol{u}_{\alpha}' \cdot \langle l,m \vert \mu  \boldsymbol{\nabla}_{\boldsymbol{w}} \delta f_\alpha \rangle 
     - \boldsymbol{F}_\alpha \cdot \langle l,m \vert \boldsymbol{\nabla}_{\boldsymbol{w}} \delta f_\alpha \rangle ,
     \\ \mathcal{B}_\alpha^{(l,m)} &= -\boldsymbol{\omega}_\alpha \cdot \langle l,m \vert \boldsymbol{w} \times   \boldsymbol{\nabla}_{\boldsymbol{w}} \delta f_\alpha \rangle,
\end{aligned}
\end{equation}
where vectors have been condensed to
\begin{equation}
\begin{aligned}
    \boldsymbol{F}_e &= \frac{1}{\rho_e}\bigg( -\boldsymbol{\nabla}  p_e -\boldsymbol{\nabla} \cdot \Pi_e  + \boldsymbol{R}_e\bigg), \quad  \boldsymbol{F}_i = \frac{1}{\rho_i}\bigg( -\boldsymbol{\nabla} p_i -\boldsymbol{\nabla}\cdot \Pi_i \bigg), \quad \boldsymbol{\omega}_e = -\frac{e}{m_e}\boldsymbol{B},\quad \boldsymbol{\omega}_i = \frac{Ze}{m_i}\boldsymbol{B}.
\end{aligned}
\end{equation}
and components
\begin{equation}
\begin{aligned}
    \langle l,m \vert \mu \delta f_e \rangle &= \frac{l+1-m}{2l+1} \delta f_e^{(l+1,m)} + \frac{l+m}{2l+1}\delta f_e^{(l-1,m)},
    \\ \langle l,m \vert \boldsymbol{\nabla}_{\boldsymbol{w}} \delta f_e \rangle &= \frac{1}{2} (\boldsymbol{\hat{x}} + i\boldsymbol{\hat{y}}) \sum_{k=0}^\infty \bigg[ a^+_{l,k}(m) \partial_w \delta f_e^{(k,m+1)} + ( -c^+_{l,k}(m) + (m+1) d^+_{l,k}(m)) \frac{1}{w}  \delta f_e^{(k,m+1)} \bigg]
    \\& + \frac{1}{2} (\boldsymbol{\hat{x}} - i\boldsymbol{\hat{y}}) \sum_{k=0}^\infty \bigg[ a^-_{l,k}(m) \partial_w \delta f_e^{(k,m-1)} + ( -c^-_{l,k}(m) - (m-1) d^-_{l,k}(m)) \frac{1}{w}  \delta f_e^{(k,m-1)} \bigg]
    \\ &+ \boldsymbol{\hat{z}} \bigg[ \frac{l+1-m}{2l+1}\frac{1}{w^{l+2}}\partial_w \Big( w^{l+2} \delta f_e^{(l+1,m)} \Big) + \frac{l+m}{2l+1}w^{l-1}\partial_w \Big( \frac{1}{w^{l-1}} \delta f_e^{(l-1,m)} \Big) \bigg],
    \\ \langle l,m \vert \mu \boldsymbol{\nabla}_{\boldsymbol{w}} \delta f_e \rangle &= \frac{1}{2} (\boldsymbol{\hat{x}} + i\boldsymbol{\hat{y}}) \sum_{k=0}^\infty \bigg[ \tilde{a}^+_{l,k}(m) \partial_w \delta f_e^{(k,m+1)} + ( -\tilde{c}^+_{l,k}(m) + (m+1) \tilde{d}^+_{l,k}(m)) \frac{1}{w}  \delta f_e^{(k,m+1)} \bigg]
    \\& + \frac{1}{2} (\boldsymbol{\hat{x}} - i\boldsymbol{\hat{y}}) \sum_{k=0}^\infty \bigg[ \tilde{a}^-_{l,k}(m) \partial_w \delta f_e^{(k,m-1)} + ( -\tilde{c}^-_{l,k}(m) - (m-1) \tilde{d}^-_{l,k}(m)) \frac{1}{w}  \delta f_e^{(k,m-1)} \bigg]
    \\ &+ \boldsymbol{\hat{z}} \sum_{k=0}^\infty \bigg[ \tilde{b}_{l,k}(m) \partial_w \delta f_e^{(k,m)} + \tilde{g}_{l,k}(m)  \frac{1}{w}  \delta f_e^{(k,m)} \bigg],
    \\ \langle l,m \vert \boldsymbol{w}\times \boldsymbol{\nabla}_{\boldsymbol{w}} \delta f_e \rangle &= \frac{1}{2i} (\boldsymbol{\hat{x}} + i\boldsymbol{\hat{y}}) \sum_{k=0}^\infty \bigg[ ( C^+_{l,k}(m) - (m+1) D^+_{l,k}(m))   \delta f_e^{(k,m+1)} \bigg]
    \\& + \frac{1}{2i} (\boldsymbol{\hat{x}} - i\boldsymbol{\hat{y}}) \sum_{k=0}^\infty \bigg[ ( -C^-_{l,k}(m) - (m-1) D^-_{l,k}(m))   \delta f_e^{(k,m-1)} \bigg]
    \\& -im \delta f^{(l,m)}_e \boldsymbol{\hat{z}},
\end{aligned}
\end{equation}
where the matrix elements follow the same definition as Swanekamp et al. \cite{10.1063/1.5109430} with the tilde coefficients indicating an additional factor of $\mu$ in the integrand of their definitions, e.g.

\begin{equation}
\begin{aligned}
    a^\pm_{l,k}(m) &\coloneqq \frac{2k+1}{2} \frac{(k-(m\pm 1))! }{(k+(m\pm 1))!} \int_{-1}^1 d\mu\ P_l^m(\mu) \sqrt{1-\mu^2} P_k^{m\pm 1}(\mu),
    \\ \tilde{a}^\pm_{l,k}(m) &\coloneqq \frac{2k+1}{2} \frac{(k-(m\pm 1))! }{(k+(m\pm 1))!} \int_{-1}^1 d\mu\ P_l^m(\mu) \mu\sqrt{1-\mu^2} P_k^{m\pm 1}(\mu).
\end{aligned}
\end{equation}

\end{strip}


\section*{Data availability}

The data that support the findings of this study are available from the corresponding author upon reasonable request. 
\vspace{0.5cm}

\bibliographystyle{iopart-num}

\bibliography{bib.bib}

\end{document}